\documentclass[useAMS,usenatbib]{mn2e}
\usepackage{txfonts}
\usepackage{natbib}
\usepackage{aas_macros}
\usepackage{graphicx}
\usepackage{subfigure}
\usepackage{mathrsfs}

\newcommand{\gsim}{\mathrel{\hbox{\rlap{\lower.55ex \hbox {$\sim$}}
                   \kern-.3em \raise.4ex \hbox{$>$}}}}
\newcommand{\lsim}{\mathrel{\hbox{\rlap{\lower.55ex \hbox {$\sim$}}
                   \kern-.3em \raise.4ex \hbox{$<$}}}}

\newcommand\sdensity{$\rm g \: cm^{-2}$}

\newcommand\rhill{$R_{\rm H}$}
\newcommand\rplanet{$R_{\rm p}$}
\newcommand\rorbit{$r_{\rm p}$}

\newcommand\mrorbit{ r_{\rm p} }

\newcommand\solarmass{$\rm M_{\odot}$}

\newcommand\earthmass{$\rm M_{\oplus}$}

\title[Planetary migration and growth]{Planet migration: self-gravitating radiation hydrodynamical models of protoplanets with surfaces}
\author[B.A. Ayliffe \& M.R. Bate]{Ben A. Ayliffe\thanks{E-mail:
ayliffe@astro.ex.ac.uk} and Matthew R. Bate\thanks{E-mail:
mbate@astro.ex.ac.uk}\\ School of Physics, University of Exeter, Stocker
Road, Exeter EX4 4QL}

\date{\today}
\begin{document}
\maketitle

\begin{abstract}

We calculate radial migration rates of protoplanets in laminar minimum mass solar nebula discs using three-dimensional self-gravitating radiation hydrodynamical (RHD) models. The protoplanets are free to migrate, whereupon their migration rates are measured. For low mass protoplanets (10-50 \earthmass) we find increases in the migration timescales of up to an order of magnitude between locally-isothermal and RHD models. In the high-mass regime the migration rates are changed very little.

These results are arrived at by calculating migration rates in locally-isothermal models, before sequentially introducing self-gravity, and radiative transfer, allowing us to isolate the effects of the additional physics. We find that using a locally-isothermal equation of state, without self-gravity, we reproduce the migration rates obtained by previous analytic and numerical models. We explore the impact of different protoplanet models, and changes to their assumed radii, upon migration. The introduction of self-gravity gives a slight reduction of the migration rates, whilst the inertial mass problem, which has been proposed for high mass protoplanets with circumplanetary discs, is reproduced. Upon introducing radiative transfer to models of low mass protoplanets ($\approx 10$ \earthmass), modelled as small radius accreting point masses, we find outward migration with a rate of approximately twice the analytic inward rate. However, when modelling such a protoplanet in a more realistic manner, with a surface which enables the formation of a deep envelope, this outward migration is not seen.


\end{abstract}

\begin{keywords}
planets and satellites: formation -- radiative transfer -- methods: numerical -- hydrodynamics -- planetary systems: formation
\end{keywords}

\section{Introduction}

The discovery of giant planets in tight orbits around their stars has introduced a complication that planet formation models must address. The formation theories contemporary to the discovery of 51 Peg b \citep{MayQue1995}, the first such exoplanet discovered, had assumed planet systems with morphologies similar to the solar system, and were unable to explain the planet's creation. The high temperatures in the inner regions of circumstellar discs, as well as the dominance of the star's gravity, prevent formation by gravitational instability. Fairing no better, the core accretion model is confounded by the low solids surface density in such high temperature regions where no ice can form. However, a mechanism that might offer an explanation for the existence of such Hot Jupiters had already been suggested by \cite{GolTre1980} who had studied the rate of angular momentum and energy transfer between a disc and an embedded planet. They found that the angular momentum exchange was such that a planet's orbital radius should change significantly on short timescales ($\sim$$10^{3} - 10^{4}$ years).

There are two primary types of planet migration which act in different mass regimes. Type I acts in the low mass regime ($\lesssim$ 100 \earthmass), and is a result of a low mass planet's interactions with annuli of gas that are in resonance with it, at so called Lindblad resonances, as well as the corotation resonance. Angular momentum exchange occurs efficiently with Lindblad resonances both inside and outside of the protoplanets orbital radius. The inner Lindblad resonance passes angular momentum to the protoplanet, acting to accelerate it, whilst the outer Lindblad resonance works in oppostion. A linear analysis can be used to describe the effect of Type I migration, and in doing so it is found that the Lindblad resonances are not spaced symmetrically inside and outside of the planet's orbit. Instead each outer Lindblad resonance is somewhat closer than the corresponding inner resonance, allowing the angular momentum exchange with these outer resonances to dominate \citep{Ward1986}. The interaction of a planet with gas in the corotation region, the so called corotation torque, has a magnitude at most of only a few tens of percent of the differential Lindblad torque \citep{Ward1997, TanTakWar2002} when fully unsaturated. As a result, the net loss of angular momentum by the planet to the disc due to the differential Lindblad torque causes the planet to migrate inwards \citep{Ward1986}.

The rates of planet migration derived from analytic models (i.e. \citealt{Ward1997, TanTakWar2002}) are in agreement with those obtained using locally-isothermal numerical models \citep{KorPol1993, NelPapMasKle2000, Masset2002, DAnHenKle2002, BatLubOglMil2003, DAnKleHen2003, AliMorBen2004, DAnBatLub2005, KlaKle2006, DAnLub2008, ParMel2008, LiLubLiLin2009}. These rates are such that low mass planets very rapidly fall into their stars, on much shorter timescales than are required for them to grow to of order a Jupiter mass. As such, to explain the observed exoplanet population, some modification is required to slow down Type I migration, something beyond what the locally-isothermal models have considered \citep{AliMorBen2004, AliMorBenWin2005, IdaLin2008, MorAliBen2009}.

One such modification is the inclusion of turbulence, dispensing with the usual laminar disc assumption. Three-dimensional magnetohydrodynamic models performed by \cite{NelPap2004} focused on the implications for migration of interactions of protoplanets within magnetorotationally unstable accretion discs. It was found that the disc turbulence led to large scale fluctuations in the net torque experienced by a protoplanet over the course of its orbit. As such the change in a protoplanet's orbital radius becomes a random-walk. The most significant fluctuations were found for protoplanets of $< 30$ \earthmass, for which the time averaged fluctuations were inconsistent with conventional Type I migration. The increase in a low mass protoplanet's migration timescale due to the oscillation of its orbit suggests turbulence as a route by which to improve the survival probability of such bodies \citep{NelPap2004, Nelson2005}. Another addition to migration models which can lead to slower rates is more realistic thermodynamics. Several numerical models have introduced non-isothermal equations of state \citep{MorTan2003, JanSas2005, PaaMel2006, PaaMel2008, BarMas2008a, ParPap2008, KleCri2008, KleBitKla2009}, and have found that the rate of Type I migration can be reduced, or its direction reversed.

The second type of migration, called Type II, becomes effective once a planet grows to a mass of a few tenths of a Jupiter mass and is able to clear its corotation region of material, forming a gap in its parent disc \citep{Ward1997,Brydenetal1999, Kley1999}. The width of the gap is determined by the competing influences of the viscous and pressure forces, which tend to close the gap, and the gravitational torques which work to open it. The planet becomes locked in the gap it has created, and migrates inwards on the disc's viscous accretion timescale. Such Type II migration continues on the viscous timescale whilst the protoplanet mass is less than or comparable to the mass of the local disc with which it is interacting. A protoplanet that comes to dominate the local mass can no longer be treated simply as a constituent part of the disc. Rather, its inertia acts to slow the rate of migration that the disc's viscous evolution can bring about \citep{SyeCla1995, IvaPopPol1999}. \citeauthor{IvaPopPol1999} suggested an expression to describe this late stage migration which led to a reduction of the migration rate with increasing protoplanet mass and with reduced orbital radius. This gives a slowing migration that might enable massive protoplanets to avoid falling into the central star, and help to form Hot Jupiters. The transition from Type I into Type II migration cannot be treated analytically due to the many non-linear processes at work, but is has been modelled numerically in both two and three-dimensions \citep{Brydenetal1999, Kley1999, DAnHenKle2002, BatLubOglMil2003}.

This paper discusses numerical models conducted to investigate planet migration, as has been done previously by numerous authors in both two and three dimensions \citep{KorPol1993, NelPapMasKle2000, Masset2002, DAnHenKle2002, BatLubOglMil2003, NelPap2003, DAnKleHen2003, AliMorBen2004, NelPap2004, Nelson2005, DAnBatLub2005, MasDAnKle2006, KlaKle2006, CriMor2007, ParPap2008, ParMel2008,DAnLub2008, KleCri2008, PepAryMel2008a, PepAryMel2008b, LiLubLiLin2009, PaaPap2009, CriBarKleMas2009, KleBitKla2009, YuLiLiLub2010, BarLin2010}. Some of the more recent of these works have begun to include more physics, such as magnetic fields, and non-isothermal equations of state as discussed above. Other models, conducted by \cite{NelPapMasKle2000}, \cite{DAnKleHen2003}, \cite{AliMorBen2004}, and \cite{AliMorBenWin2005}, have included mass and angular momentum accretion onto migrating protoplanets, whilst the influence of self-gravity was explored by \cite{NelBen2003}, \cite{BarMas2008}, and analytically by \cite{PieHur2005}. Regardless of the physics included, it has been common to exclude torques from the immediate vicinity of the planet due to the difficulty of resolving this region in any great detail. This poor resolution also prevents the self-consistent modelling of a protoplanetary envelope, which requires both good resolution and the inclusion of self-gravity and radiative transfer.

In this work we conduct three dimensional global disc models, using smoothed particle hydrodynamics (SPH), that include radiative transfer, self-gravity, and which use a planetary surface to allow modelling of gas flow to well within the Hill sphere. With our surface treatment, and the adaptable resolution of SPH, the innermost torques can be well resolved and a realistic protoplanetary envelope can develop. We conduct models of varying complexity, introducing more physics in stages such that we can determine the impact of each addition on protoplanet migration. The first models we discuss are designed to emulate the treatment used in the grid models of \cite{BatLubOglMil2003}, allowing us to test our SPH calculations. We then introduce mass and angular momentum accretion, followed by the addition of self-gravity. Next the surface treatment is employed to explore the impact of realistic gas flow within the Hill sphere. Finally radiative transfer is folded into the mix, and a range of opacities explored.
 
In Section 2, we describe our computational method and its testing. In Section 3 we present the results of our models, starting with the simplest locally-isothermal case, and progressing to self-gravitating radiation hydrodynamic cases using our planet surface treatment. Section 4 discusses the implications of our results for giant planet formation theory, whilst a summary and our conclusions are given in Section 5.

\section{Computational Method}
\label{sec:setup}

The calculations described herein have been performed using a three-dimensional SPH code. This SPH code has its origins in a version first developed by \citeauthor{Benz1990} (\citeyear{Benz1990}; \citealt{BenzCamPreBow1990}) but it has undergone substantial modification in subsequent years. Energy and entropy are conserved to timestepping accuracy by use of the variable smoothing length formalism of \cite{SprHer2002} and \cite{Monaghan2002} with our specific implementation being described in \cite{PriBat2007}. Gravitational forces are calculated and neighbouring particles are found using a binary tree. Radiative transfer is modelled in the two temperature (gas, $T_{\rm g}$, and radiation, $T_{\rm r}$) flux-limited diffusion approximation using the method developed by \citet{WhiBatMon2005} and \citet{WhiBat2006}.  Integration of the SPH equations is achieved using a second-order Runge-Kutta-Fehlberg integrator with particles having individual timesteps \citep*{Bate1995}. The code has been parallelised by M. Bate using OpenMP.

\subsection{Disc setup}

We model a protoplanetary disc ($2\pi$ radians) with radial bounds of 0.1 - 3 \rorbit \ (0.52 - 15.6 AU), where \rorbit \ is the initial orbital radius of our embedded protoplanets, taking a value of 5.2 AU. The radial temperature profile for the disc is initialised as $T_{\rm g} \propto r^{-1}$, whilst the surface density of the disc goes as $\Sigma \propto r^{-1/2}$; these profiles are equivalent to those of \cite{BatLubOglMil2003}, giving a constant ratio of disc scaleheight to radius of $H/R \approx 0.05$. At \rorbit \ the initial temperature is $\approx$ 75K and the surface density is 75 \sdensity. This leads to a total disc mass of $\approx 0.005$ \solarmass \ within the boundaries, (or $\approx 0.0075$ \solarmass \ within 1.56 - 20.8 AU inline with \citealt{BatLubOglMil2003}, or 0.01 \solarmass \ within 26 AU inline with \citealt{DAnKleHen2003}). The disc begins with a Keplerian velocity structure, established around a 1 \solarmass \ star. It should be noted that when modelled with radiation hydrodynamics the disc settles to have a slightly steeper temperature profile at the midplane, where in our disc it becomes approximately $T_{\rm g} \propto r^{-1.3}$. A similar steepening is found by \cite{KleBitKla2009} in their radiation hydrodynamical models.

The inner edge of the disc is bounded by a potential barrier which acts to prevent the disc accreting on to the central star. The outer edge of the disc is surrounded by a population of ghost particles that act as a pressure barrier to prevent the disc from spreading due to shear viscosity. Particles that do move beyond the outer edge of the computational domain are removed from the calculation. The migration rates of planets are determined by the structure of the disc in their vicinity, particularly the surface density profile. As such, the aim of these boundaries was to establish a broad region of disc about \rorbit \ throughout which the surface density followed the desired profile, with the only modification being due to the planet-disc interaction. Maintaining a broad stable region is important to allow any embedded protoplanet to migrate away from its initial position at \rorbit, and to allow density waves to propagate to large distances, such that their interactions with the protoplanet are included to the point of inconsequence.

\begin{figure}
\centering
\includegraphics[width=8cm]{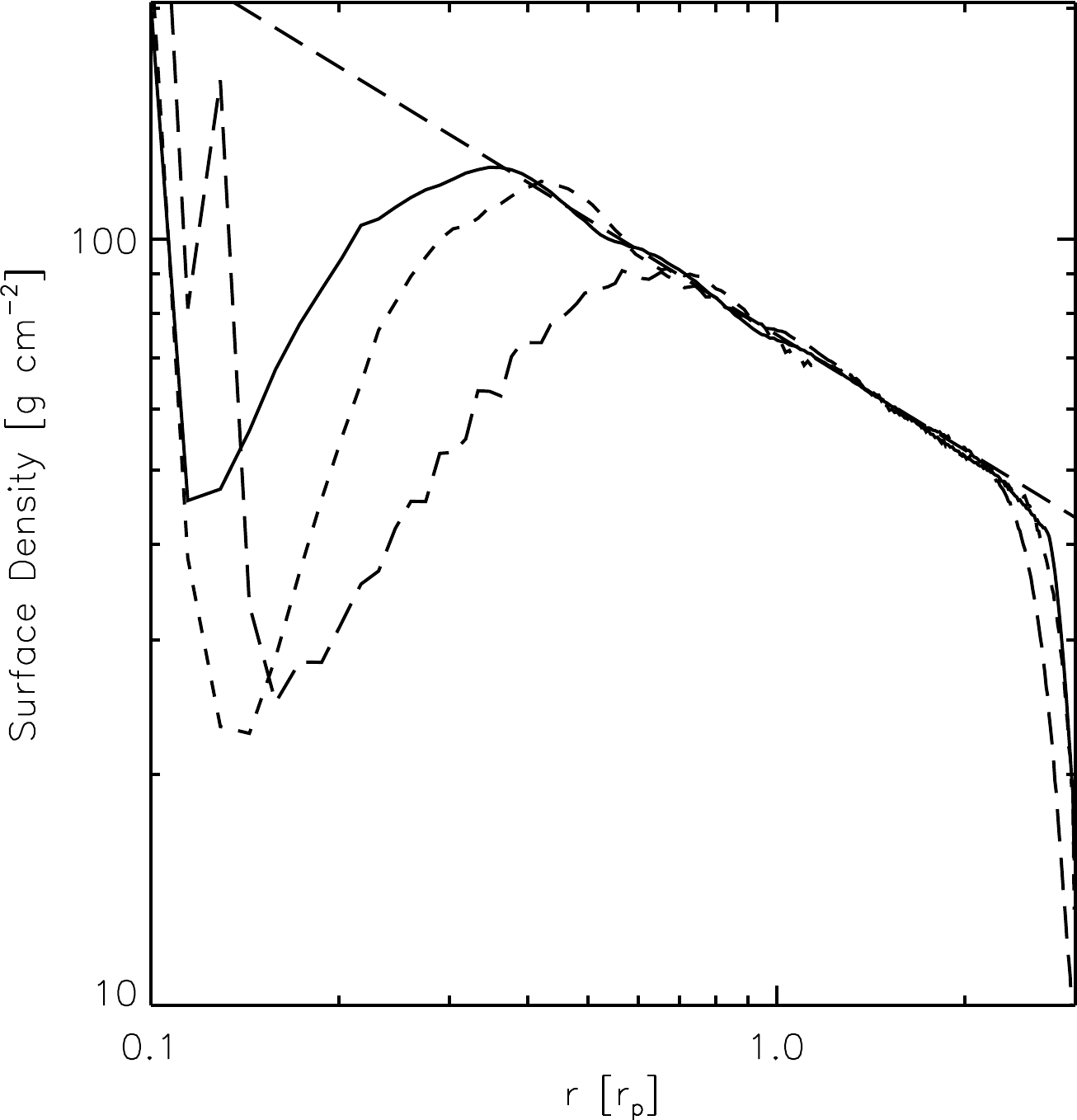}
\caption{Surface density profiles of a locally-isothermal disc modelled with $10^{7}$ (solid line), $2 \times 10^{6}$ (short-dashed line), and $5 \times 10^{5}$ (long-dashed line) particles, in the absence of a planet, after 4 orbits of the disc's outer-edge. The straight dashed line across the top denotes the desired surface density profile of $\Sigma(r) \propto r^{- \frac{1}{2}}$, with a value of 75 \sdensity \ at $r = \mrorbit$. With the implemented disc boundaries this profile is achieved by the two higher resolution calculations in the region over which a planet may migrate, whilst in the lowest resolution calculation the effect of the inner boundary reaches to almost \rorbit.}
\label{fig:miginitial}
\end{figure}

The inner boundary allowed gas to pile up at a small radius, far from the intended initial orbital radius of any introduced protoplanets. Once gas has accumulated at the boundary, the concentration of material exerts an outward pressure gradient back into the disc, supporting further gas against inward migration. Fig.~\ref{fig:miginitial} illustrates the disc profile that develops, and illustrates the success of the boundaries in preserving the desired surface density profile (shown by the long-dashed line) throughout the region of likely migration, from 0.5 to 2 \rorbit \ (2.6 - 10.4 AU). The profiles shown are for calculations with $5 \times 10^5$, $2 \times 10^{6}$ and $10^{7}$ particles. In the highest resolution calculation, the rarefaction due to the outwards pressure from the boundary is narrowest and shallowest. However, the radial range over which a consistent surface density profile is achieved is not greatly changed between the two higher resolution models. For calculations performed with $5 \times 10^{5}$ particles this is not the case, with a much greater encroachment of the boundary effects into the likely migration region. In the calculations that follow we use $2 \times 10^{6}$ particles as this was sufficient to establish a stable disc in the planet's vicinity, whilst being considerably faster than using $10^{7}$ particles, for which the benefits were marginal.

Upon creating a disc model, there is generally a period of transient structural perturbations as the model settles to a stable state. This settling must be completed before protoplanets are introduced in order that their orbits are not immediately disturbed by perturbations unrelated to the interesting protoplanet-disc interactions. The initial discs were evolved in the absence of a planet until any transience resulting from settling had dissipated, which required just over 4 orbits of the disc's outer edge.

\subsection{Equation of state}

Two equations of state are used for the calculations presented in this paper.  One is a locally-isothermal equation of state with the temperature of the gas throughout the disc being a fixed function of radius (as described in the disc setup). The second is used for our radiation hydrodynamical calculations, which are conducted using an ideal gas equation of state $p=\rho T_{g} R_{g}/\mu$ where $R_{g}$ is the gas constant, $\rho$ is the density, $T_{g}$ is the gas temperature, and $\mu$ is the mean molecular mass. The equation of state takes into account the translational, rotational, and vibrational degrees of freedom of molecular hydrogen (assuming a 3:1 mix of ortho- and para-hydrogen; see \citealt{BolHarDurMic2007}). It also includes the dissociation of molecular hydrogen, and the ionisations of hydrogen and helium.  The hydrogen and helium mass fractions are $X=0.70$ and $Y=0.28$, respectively, whilst the contribution of metals to the equation of state is neglected.  More details on the implementation of the equation of state can be found in \cite{WhiBat2006}.

The two temperature (gas and radiation) radiative transfer in the flux-limited diffusion approximation employed in this work is implemented as described by \cite{WhiBatMon2005} and \cite{WhiBat2006}. Briefly, work and artificial viscosity (which includes both bulk and shear components) increase the thermal energy of the gas, and work done on the radiation field increases the radiative energy which can be transported via flux-limited diffusion. The energy transfer between the gas and radiation fields is dependent upon their relative temperatures, the gas density, and the opacity, $\kappa$.

\subsection{Opacity treatment}

We use interpolation of the opacity tables of \cite{PolMcKChr1985} to provide the interstellar grain opacities for solar metallicity molecular gas, whilst at higher temperatures when the grains have sublimated we use the tables of \cite{Alexander1975} (the IVa King model) to provide the gas opacities (for further details see \citealt{WhiBat2006}). The grain opacities are based on interstellar abundances, and may not hold true in a circumstellar disc where agglomeration can strongly modify the grain population, reducing the opacity. Therefore, we vary the grain opacity in our models to mimic the effects of these processes; the gas opacities of \cite{Alexander1975} that are used at higher temperatures are not scaled. In this paper we use 100, 10, and 1 per cent interstellar grain opacities (see \citealt{AylBat2009} for further details).

\subsection{Radiation boundary}

Many of the calculations performed here involved radiative transfer, and so required a mechanism by which energy could be lost from the disc into the encompassing vacuum. This necessity arises because the flux-limited diffusion scheme transfers energy between SPH particles, rendering it unable to radiate into a vacuum where there are no particles. By determining the height above and below the midplane at which the optical depth into the disc, $\tau_{\rm op} \approx 1$, we can identify two layers of particles which exist in optically thin regions. Compelling particles that fall within these regions to follow the initial temperature profile of the disc allows the energy they receive from the bulk of the disc to be effectively lost, as though into the vacuum. This initial temperature profile was chosen for purposes of comparison with previous work where it is common, and crudely represents the insolation of the disc.



The vertical density structure of the disc approximately follows a Gaussian distribution, integration over which yields the surface density, $\Sigma_{r}$, in terms of the error function, erf(). The optical depth is given by $\tau_{\rm op}(a) = - \int_{a}^{\infty} \kappa \rho dz$. Completing this integration with the Gaussian form of $\rho$, setting $\tau_{\rm op} = 1$, and the upper limit of integration set to infinity to represent the vacuum, it is possible to rearrange to find the lower limit of integration, depth `$a$', as shown in equation \ref{eq:radiationboundary}.

\begin{equation}
\frac{a}{H} = \sqrt{2} \, {\rm erf^{-1}} \left(1 - \frac{2 \tau_{\rm op}}{\kappa \Sigma_{r}} \right).
\label{eq:radiationboundary}
\end{equation}

\noindent This is the depth in the disc from above which radiation can escape to the vacuum, and so defines the height of our radiation boundary. The boundary height varies with radius, due to the radial dependence of $\Sigma_{r}$ and $\kappa$, the values of which are taken from the initial disc distribution.

\subsection{Planet treatments}

Once a suitable disc has been established, it is possible to embed a protoplanet and model its evolution. In all of the models discussed in this paper, the protoplanet is free to move from its initial orbital radius, and its migration rate is measured, rather than calculated from static torques. How the protoplanet interacts with the disc, and so migrates, depends upon the way in which the protoplanet is modelled, and we employ several methods in this work. In the first case gas is not accreted, but instead gas that comes within a specified radius, $r_{\rm acc}$, of the point mass representing the protoplanet, is removed from the calculation, and its energy, momentum, and mass are all dispensed with. We call the particles used for this type of protoplanet treatment Killing sink particles, and they migrate purely due to torque interactions with the disc, approximately recreating the treatment used in \cite{BatLubOglMil2003}. Typically we set $r_{\rm acc}$ to be some fraction of the Hill radius, \rhill.

The second treatment also models the protoplanet as a sink particle which removes gas that comes within $r_{\rm acc}$, but in this case the gas properties are added to the sink particle. These sink particles are called Accreting sinks. They are able to absorb the energy, mass, linear momentum, and angular momentum of the accreted gas \citep[see][]{BatBonPri1995}.

The third and final protoplanet treatment is one that we have used in previous work \citep{AylBat2009,AylBat2009b} which uses a surface force that allows gas to pile up on the surface of a protoplanet, realistically modelling accretion. This scheme is now applied by wrapping the surface around a gravitating point mass that is free to move, and which represents the protoplanet's initial mass. Such a protoplanet can move in response to changes in the momentum of the point mass and its bound envelope, the extent of which is self-consistently arrived at, rather than being prescribed. The surface is applied as a modification to the usual gravitational force in the protoplanet's immediate vicinity, taking the form

\begin{equation}
F_{r} = - \frac{GM_{p}}{r^{2}}\left(1 - \left(\frac{2R_{p}-r}{R_{p}}\right)^{4}\right),
\label{eq:surface}
\end{equation}

\noindent for $r <2$ \rplanet, where \rplanet \ is the protoplanet radius, $r$ is the radius from the centre of the protoplanet, and $M_{\rm p}$ is the protoplanet's initial mass.  This equation yields zero net force between a gas particle and the gravitating point mass at the surface radius $R_{\rm p}$. Inside of this radius the force is outwards and increases rapidly with decreasing radius, such that the inner most gas particles come to rest very close to the surface, with an equilibrium position slightly inside due to the envelope weight pressing down. As in our previous work, we provide a smooth start to the calculations by embedding a protoplanet of radius $R_{\rm p} =0.01 r_{\rm p}$ which then shrinks exponentially to the desired radius during the first orbit of the protoplanet.

\subsection{Resolution dependence of migration}

\begin{figure}
\centering
\includegraphics[width=1.0 \columnwidth]{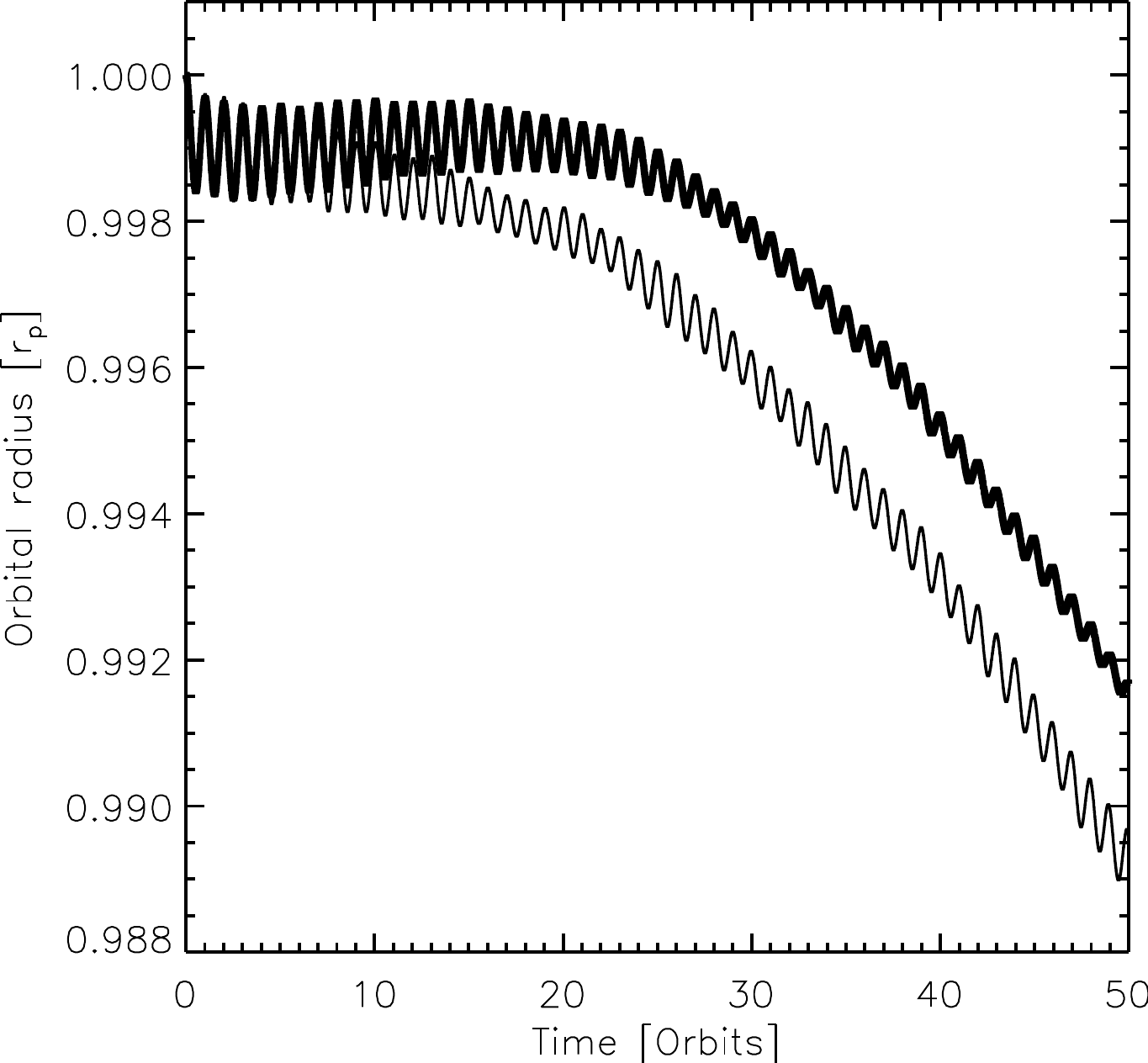}
\caption{The orbital evolution of a 10 \earthmass \ protoplanet modelled using Accreting sink particles, with accretion radii of $\sim$0.2 \rhill \ (a radius chosen to mimic the ZEUS models of \citealt{BatLubOglMil2003}). Two resolutions were used to perform otherwise identical calculations, $10^{7}$ particles (thick upper line) and $2 \times 10^{6}$ particles (thin lower line). The evolution covers 50 orbits of the protoplanet using a locally-isothermal equation of state. The model at both resolutions establishes the same rate of migration after a period of settling, roughly equal to the libration time.}
\label{fig:migrescomp}
\end{figure}

Resolution was considered previously in relation to the disc boundaries, and establishing the desired density profile in an undisturbed disc. However, ensuring that the planet-disc interaction is well resolved in our models, and that the migration rates obtained are independent of resolution, required further testing. To this end, two simulations were performed which were identical in every respect excepting their resolution. A 10 \earthmass \ protoplanet, modelled with an Accreting sink particle, was used for these calculations. The spiral waves launched by such a low mass planet are weak, as are the resulting torques acting back upon the planet. So whilst a degree of migration is expected, insufficient resolution might poorly resolve these small effects; this makes such a planet an ideal candidate for testing resolution. Fig.~\ref{fig:migrescomp} shows the evolution of the planet's orbital radius with time for two resolutions.

There is an initial period of settling whilst the torques due to the horseshoe region are established. As was seen in \cite{Masset2002}, the outward corotation torques almost balance the inward differential Lindblad torques when the protoplanet is first embedded leading to almost no migration. However, as the model evolves the corotation torques become partially saturated, reducing their magnitude and allowing the Lindblad torques to dominate. As a result the protoplanet's migration is seen to accelerate until the corotation torques reach a quasi-equilibrium state, which takes of order a libration time \citep{KleBitKla2009}, given by

\begin{equation}
\tau_{lib} = \frac{4}{3} \frac{\mrorbit}{x_{s}} P_{\rm p},
\end{equation}

\noindent where $x_{s}$ is the width of the horseshoe region, and $P_{\rm p}$ is the planet's orbital period (11.8 years in the calculations presented here). Following \citeauthor{KleBitKla2009}, by using an approximation for the width of the horseshoe region, taken from the locally-isothermal models of \cite{MasDAnKle2006}, it is possible to get an idea of the horseshoe settling time. For a 10 \earthmass \ protoplanet $\tau_{lib} \sim 50$ orbits, whilst for a 333 \earthmass \ protoplanet $\tau_{lib} \sim 8$. It appears from Fig.~\ref{fig:migrescomp} that a clear migration trend is established in around 30 orbits. Once Type I migration is established, the migration rate is consistent between the two different resolution calculations. It is not clear why the initial settling appears to be resolution dependent, but the agreement of the eventual rates gives us confidence in the migration rates obtained using $2 \times 10^{6}$ particles.

\subsection{Viscosity}
\label{sec:viscosity}

Viscosity is the means by which angular momentum is transported through a laminar circumstellar disc, enabling material to flow in towards the star. The viscosity envisaged is not caused by a molecular interaction, which is insufficient to explain the rates of accretion inferred from observations. The primary alternative is thought to be a magneto-rotational instability \citep{BalHaw1991}.

Within the models presented here the viscosity is not achieved by modelling the responsible processes, but instead by using a parameterised form of viscosity, though somewhat different to a typical Shakura-Sunyaev $\alpha$-viscosity \citep{ShaSun1973}. An artificial viscosity for SPH, designed to conserve linear and angular momentum, was introduced by \cite{MonGin1983}, and modified by \cite{Monaghan1992} to deal with high mach number shocks. This artificial viscosity is not designed to represent real viscosity, but instead to allow the modelling of shock phenomena, and to damp numerical noise. The $\alpha$ and $\beta$ terms set the strength of the viscosity, with typical values of $\alpha = 1$ and $\beta = 2$. The two terms serve different functions. The $\alpha$-term establishes a viscosity to damp subsonic velocity oscillations that may be produced in the wake of a shock front. The $\beta$-term prevents particle interpenetration in supersonic shocks.

The viscosity is only applied when the gas is under compression, ideally near shocks, and we use the switch developed by \cite{MorMon1997} to try and reduce the action of artificial viscosity where the cause is not a shock. Using this switch means that instead of the global $\alpha$ being applied to all the particles, they each scale this value (down to a minimum value of $\alpha_{\rm min}$) based on their proximity to a shock. Setting $\alpha_{\rm min} = 0.1$ can significantly reduce unwanted dissipation. The differential rate of rotation of the disc leads to a shear viscosity by which angular momentum is exchanged through the disc. The value of the viscosity at any point can be stated as a Shakura-Sunyaev $\alpha$-viscosity, $\alpha_{\rm SS}$, using

\begin{equation}
\alpha_{\rm SS} \simeq 0.05 \alpha \frac{h}{H}
\end{equation}

\noindent as given by \cite{LodPri2010}, where the coefficient given here is a factor 2 smaller because the viscous force is only applied between approaching, not receding, particles. Here $h$ is the local average SPH smoothing length, and $H$ is the disc scaleheight. Around \rorbit \ these quantities have values of approximately 0.016, and 0.05 respectively, with $\bar{\alpha} \approx 0.25$ giving  $\alpha_{\rm SS} \approx 0.004$. This is typical for such a disc, and inline with the models of \cite{BatLubOglMil2003} upon which our initial conditions are based.

\subsection{Measuring the migration rate}

\begin{figure}
\centering
\includegraphics[width=1.0 \columnwidth]{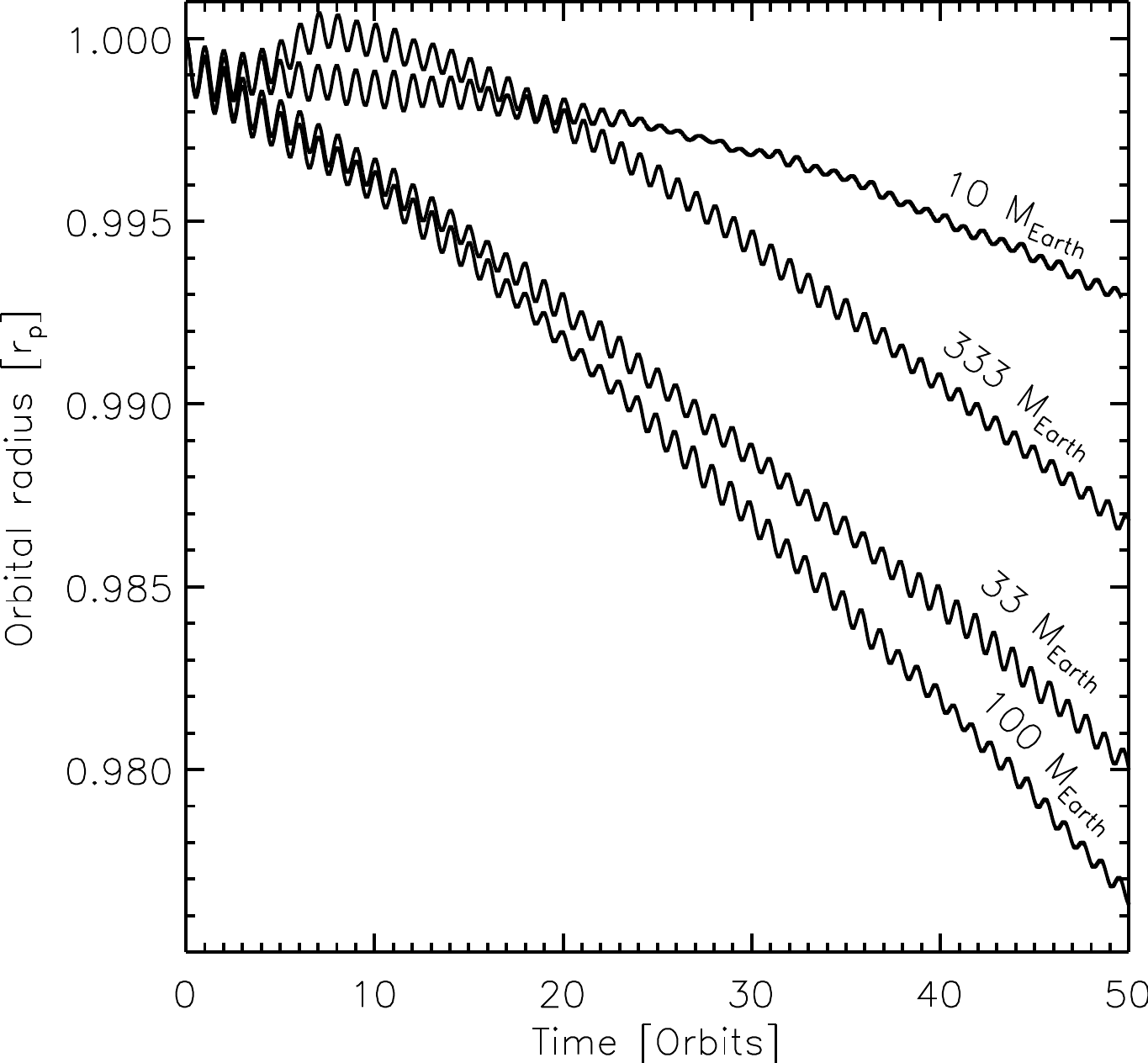}
\caption{The orbital evolution of 4 protoplanets modelled in locally-isothermal discs, using Accreting sink particles with $r_{\rm acc} = 0.1$ \rhill. In all instances the planetary migration is well underway and free from large scale fluctuations after 25 orbits. The migration rates are measured over the last 25 orbits. The Jupiter mass protoplanet appears to settle in $\approx 10$ orbits, inline with the libration time of $\approx 8$ orbits \protect \citep{KleBitKla2009}. }
\label{fig:libration}
\end{figure}

The migration rates were calculated by taking linear fits of the orbital radii evolution data (e.g. Fig.~\ref{fig:migrescomp}). Each simulation was allowed to evolve for at least 50 orbital periods, (as was done in \citealt{BatLubOglMil2003}), of which only the latter 25 were used for fitting purposes. The early orbits are sacrificed to ensure that any transient disruption to the disc caused by the sudden introduction of a planet have subsided and to allow gas in the horseshoe region to reach a quasi-equilibrium state, as discussed previously. Fig.~\ref{fig:libration} illustrates the orbital migration of 4 different mass protoplanets modelled with Accreting sinks and a locally-isothermal equation of state. It can be seen that the final 25 orbits of evolution are settled, and so provide a suitable period over which to measure the migration rates. For the highest mass protoplanets it can be seen that their migration over the last 25 orbits is linear in time, suggesting that the rate is well established. Deviations from linearity, which might cause a long term change in the rate are very subtle, and would only be established on very long timescales, inaccessible to our models due to their computational expense.

There are instances where measuring the migration rate over the last 25 orbits will lead to faster rates for a stated mass than should truly be ascribed to it. In the first 25 orbits, in the Accreting sink and point mass with surface calculations, the protoplanet's mass has increased through accretion. As a result of this, the migration rate tends to accelerate, giving a non-linear orbital radius evolution. It is therefore the case that linear fits to the latter half of this evolution don't exactly render the gradient expected for the initial protoplanetary mass. The impact of such mass accretion most strongly influences the orbital evolution of the lowest mass cores, which accrete a higher fraction of their total mass. As a result, we present the migration rates with uncertainties in the mass, portrayed by error bars which stretch across the mass range over which the linear fit is taken, with the migration timescale plotted at the mean mass. In the Accreting sink calculations, the mass of the protoplanet includes that of all the gas accreted, but for the point mass with surface calculations the accreted mass is measured as the mass within the Hill radius. For a majority of the calculations, the mass spread is very small, and thus the mass associated with a given migration rate is well known. The migration timescale, $\tau$, is then calculated as the time a planet of fixed mass would take to migrate from its starting orbital radius, (5.2 AU in all of the models discussed), in to the central star; $\tau = r_{\rm p}/\dot{r}$. The rate of migration is assumed constant in calculating these timescales, allowing ready comparison with previous works.

\section{Results}
\label{mig:results}

\subsection{Killing sinks}
\label{sec:migmodelcomp}

\begin{figure}
\centering
\includegraphics[width=1.0 \columnwidth]{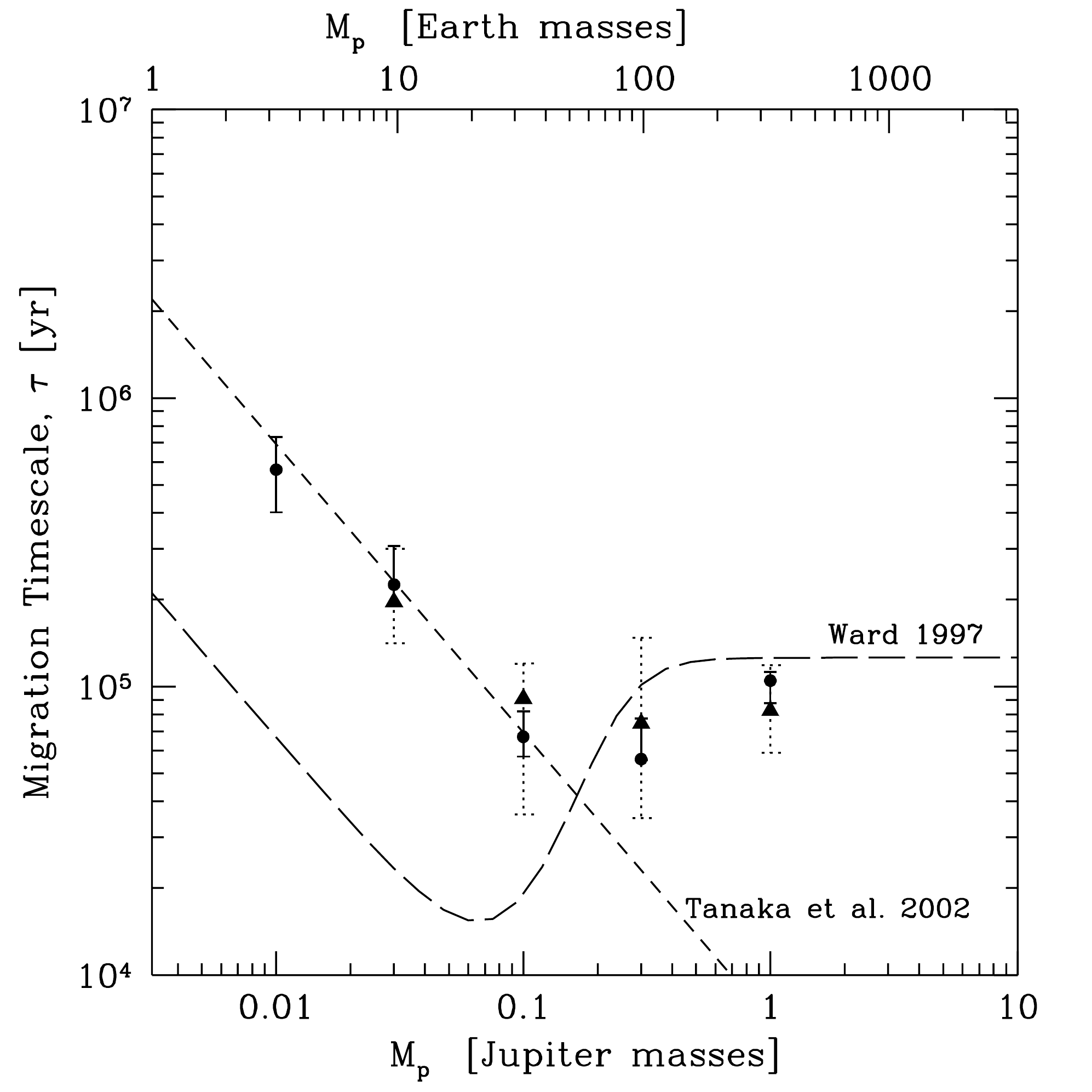}
\caption{A reproduction of figure 10 from \protect \cite{BatLubOglMil2003} showing the migration rates they obtained by considering the torques at work upon protoplanets of fixed orbital radius in their models. The dots (with solid error bars) denote the rates calculated when the torques outside of \rhill \ were included, the lower error bars are the rates when including torques external to 0.5 \rhill, and the upper error bars are the rates for just the torques beyond 1.5 \rhill. The analytic Type I model of \protect \cite{TanTakWar2002} and the Type I/II model of \protect \cite{Ward1997} are plotted as short-dashed and long-dashed lines respectively. The triangles (with dotted error bars) mark the measured migration timescales from the SPH models in which the protoplanet is able to migrate; the Killing sink particle accretion radii are \rhill. The lower and upper error bars for the SPH cases mark the measured rates with Killing sink particles of 0.5 and 1.5 \rhill \ accretion radii respectively. The SPH models show good agreement with the grid-based ZEUS models.}
\label{fig:zeuscomp}
\end{figure}

\begin{figure*}
\centering
\subfigure 
{
    \includegraphics[width=1.0 \columnwidth]{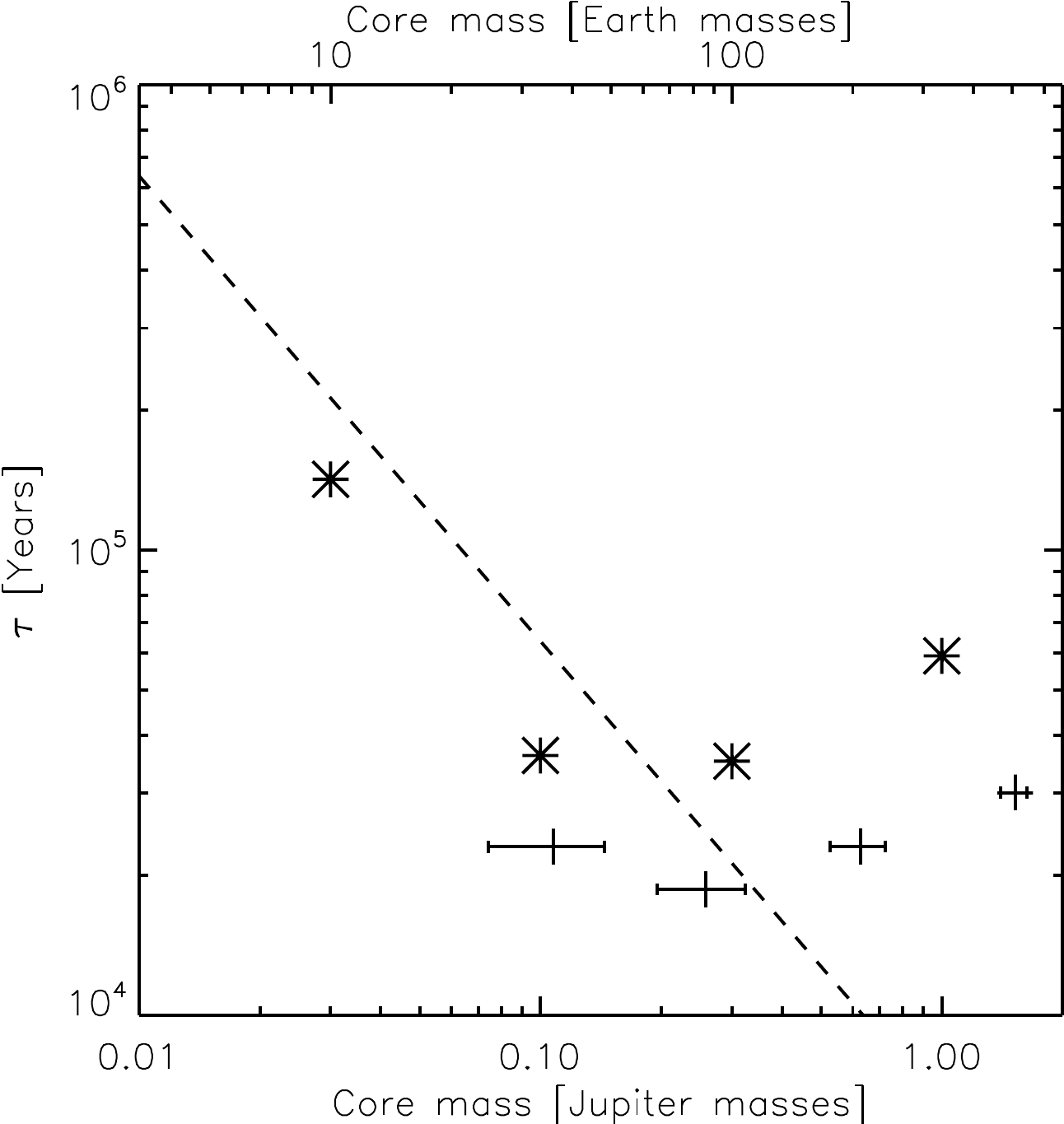}
}
\subfigure 
{
    \includegraphics[width=1.0 \columnwidth]{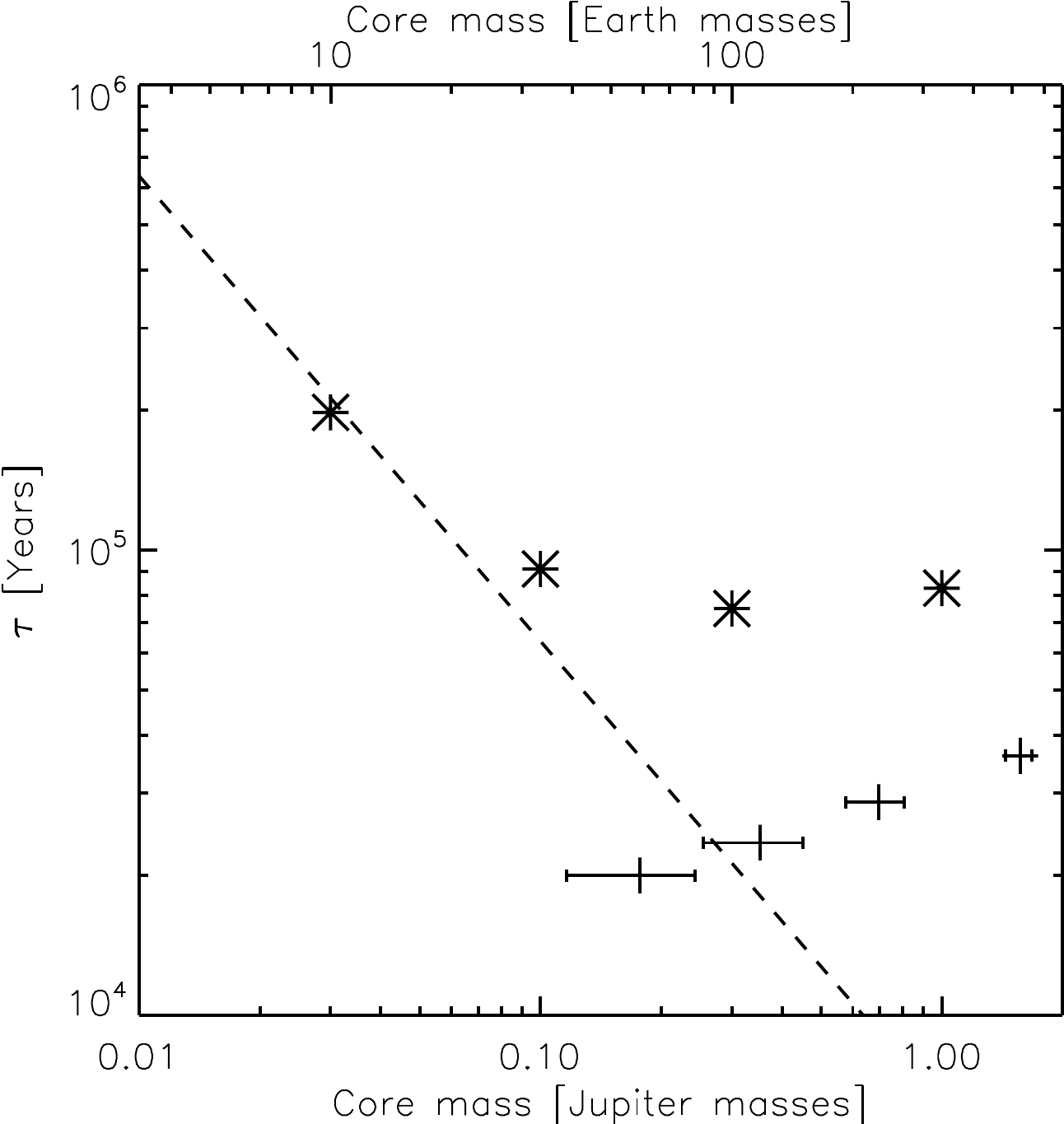}
}
\caption{A comparison of the protoplanet migration timescales measured in locally-isothermal, non-self-gravitating models using Killing sinks (asterisks), that exclude mass and angular momentum accretion, and otherwise identical calculations using Accreting sink particles (plus symbols) which include such accretion. The left panel is for those calculations where $r_{\rm acc} = 0.5$ \rhill, whilst in the right panel $r_{\rm acc} = 1.0$ \rhill. The analytic Type I model of \protect \cite{TanTakWar2002} is plotted as a dashed line. Note that the Accreting sinks begin with the same four masses as the Killing sinks, but their masses increase substantially during the calculations, particularly the low-mass protoplanets. The mass error bars for the Accreting sink particles show the range in mass over which the linear fit to determine the migration rate was applied. Including mass and angular momentum accretion consistently leads to faster rates of migration when measured at equivalent masses.}
\label{fig:typebvsa}
\end{figure*}

We began by demonstrating that out SPH code can reproduce the migration rates obtained from past studies. These comparisons are between the migration rates obtained using our Killing sinks in locally-isothermal discs, and those obtained in ZEUS finite difference models by \cite{BatLubOglMil2003}, who used the same initial conditions. The Killing sinks are used to approximately imitate the planet treatment of \citeauthor{BatLubOglMil2003}. We expect our SPH models to reproduce similar migration timescales to these previous grid based results, however, \citeauthor{BatLubOglMil2003} calculate their migration timescales by examining the net torque due to gas outside the planet's Hill radius upon the planet, which was on a fixed orbit at 5.2 AU. In making this comparison, the implicit assumption is that the torques do not change significantly when the planet is allowed to migrate rather than being held on a fixed orbit. Indeed, over the 50 orbits that are modelled in the SPH calculations, the planets do not migrate any considerable distance, as can be seen in Fig.~\ref{fig:libration}. A reproduction of Fig.~10 from \cite{BatLubOglMil2003} is shown in Fig.~\ref{fig:zeuscomp}, where the ZEUS results are plotted with circles and solid error bars, whilst the SPH derived migration timescales are plotted using triangles with dotted error bars. The term error bar is a misnomer, instead the extremes of these bars indicate the migration timescales calculated by including more (0.5 \rhill, lower bound) or less (1.5 \rhill, upper bound) of the gas closest to the protoplanet in the torque calculations. In the SPH models these same limits are found by conducting simulations with killing radii of $r_{\rm acc} = 0.5$ \rhill, 1 \rhill, and 1.5 \rhill; note that this is not exactly equivalent to changing the torque exclusion radius, and the limitations of the comparison will be discussed in the following paragraphs. For all protoplanet masses the migration timescales obtained with the SPH code, using killing radii of \rhill, differ by less than 35 per cent from the ZEUS based results. The SPH results also show the expected transition from the linear Type I migration regime to the non-linear Type II beyond $\sim 0.1{\rm M_{J}}$. As a result, we are confident that our SPH code is capable of modelling migration proficiently.


The spread in the migration timescales we find for the various killing radii are larger that those found in the ZEUS calculations with different torque exclusion radii. Explanations for these larger spreads can be found in the density and resolutions of the two models. For a given mass protoplanet, there is only one ZEUS model upon which the three different torque exclusion radius calculations are based. This gives a consistent density structure between the different cases, but a fixed spatial resolution which is poor near to 0.5 \rhill. In the SPH calculations, the $r_{\rm acc}$ equivalent to these exclusion radii are each probed in separate models. The evacuated region for an $r_{\rm acc} = 1.5$ \rhill \ Killing sink is far larger than for the 0.5 \rhill \ case, which changes the circumplanetary density structure between the two cases, with lower densities at equivalent radii for the larger $r_{\rm acc}$. This leads to lower torques acting from the boundary of a Killing sink with $r_{\rm acc} = 1.5$ \rhill \ than at an equivalent exclusion radius in the ZEUS models; hence the larger migration timescale. 

For $r_{\rm acc} = 0.5$ \rhill \ the SPH resolution is far better than that of the ZEUS calculations, and so the models better discern the gas that is available to exert torques, leading to faster migration. At all the radii, regardless of the different spreads, there is consistently a turn away from Type I to Type II migration at higher masses, and the absolute migration timescales are within a factor of two or better at each mass and radius. The figure makes clear that in both the SPH and ZEUS models, the migration timescales reduce when the killing radius (or torque exclusion radius) is shrunk. This indicates that the contributions from torques from within \rhill \ are negative, as was found in \cite{BatLubOglMil2003}. This is an important agreement as the region within \rhill \ is not treated in analytic descriptions of migration, and so is not well understood. It appears that the persistent torques from within \rhill \ are negative, which is in the opposite sense to the corotation torques in a disc with the surface density profile modelled here. We return to this in Section \ref{sec:isosurfaces}.

\subsection{Accreting sinks}
\label{sec:accsinks}


Whilst Killing sinks provided a means by which to test and compare our models, they do not allow us to include protoplanet growth. To that end we next introduced accretion into the migration models, using our Accreting sinks. Fig.~\ref{fig:typebvsa} compares the migration timescale found for Accreting sink particles in locally-isothermal models with those obtained using the Killing sink method previously described. Whilst the Killing sinks maintain their initial masses throughout, the Accreting sink models lead to increasingly massive protoplanets, sometimes growing to several times their initial mass. This makes interpreting the relative migration rates somewhat difficult, though the broad result is that including accretion leads to faster migration rates. Comparing an Accreting sink which has grown from an initial mass of 10 \earthmass \ to beyond 33 \earthmass \ with a Killing sink which maintains a mass of 33 \earthmass \ throughout its evolution must be done with some care as there are several factors contributing to their differing migration timescales. One factor is that the Accreting sink has a smaller $r_{\rm acc}$ than the 33 \earthmass Killing sink model  as the former is calculated for its initial mass of 10 \earthmass, and so it develops a different circumplanetary structure. As was discussed in section \ref{sec:migmodelcomp} for the different sized Killing sinks, a smaller $r_{\rm acc}$ leads to faster migration. This factor therefore contributes some fraction of the difference in migration rates seen in Fig.~\ref{fig:typebvsa}.

As seen in Fig.~\ref{fig:zeuscomp}, and again in comparing the left and right panels of Fig.~\ref{fig:typebvsa}, increasing $r_{\rm acc}$ reduces the migration rates of the Killing sinks. However, the Accreting sink migration rates are largely unaffected by the change in $r_{\rm acc}$ from 0.5 to 1~\rhill. The torques experienced by these sinks are equivalent to those of the Killing sinks. It therefore must be the case that the loss of negative torques from within \rhill \ as $r_{\rm acc}$ is increased is countered in the case of Accreting sinks by greater mass accretion. To test this we performed a series of calculations in which Accreting sinks were embedded in discs with which they did not gravitationally interact. These sinks accrete only material that they sweep up from their path through the disc as they cannot gravitationally draw in gas. This also means that they experience no torques from the disc. Any migration of these protoplanets is a result solely of the dilution of their specific angular momentum due to accretion of pressure supported, and so sub-Keplerian, gas.

To test this we numerically integrate from a protoplanet's initial mass and radius to its final mass, accounting for its accretion of gas with sub-Keplerian angular momentum to compare its final orbital radius with that obtained from the SPH models. In our unperturbed disc gas, at the initial orbital radius of the protoplanet, the specific angular momentum of the gas has a value of 0.998 of the protoplanet's specific angular momentum. However, we note from the SPH models that the mean radius from which gas is accreted is smaller than \rorbit, and so the average specific angular momentum is somewhat smaller than this value. Employing this measured value in our simple integration we obtain final protoplanet orbital radii that are in good agreement with those measured from the no-interaction SPH models. Applying this iterative calculation to the Accreting sink models shown in Fig.~\ref{fig:typebvsa} can account for a surprisingly large fraction of the change in their orbital radii. For a 10 \earthmass \ protoplanet, about 50\% of the change can be attributed to the effects of accreting sub-Keplerian material. This implies up to a factor of 2 difference in the migration timescale between an Accreting sink and a Killing sink due to gas accretion.

\begin{figure}
\centering
\includegraphics[width=1.0 \columnwidth]{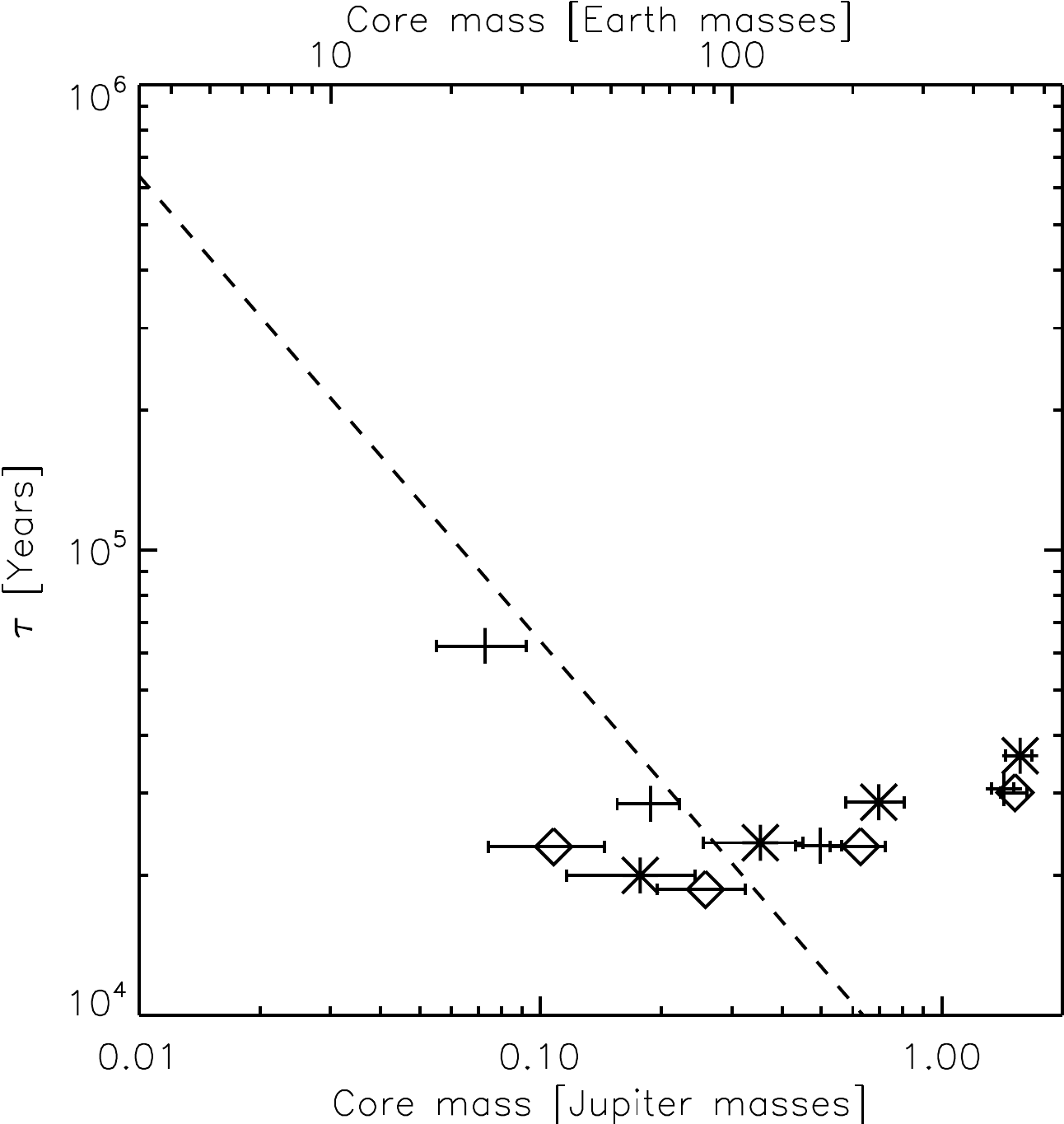}
\caption{Migration timescales for 10, 33, 100, and 333 \earthmass \ protoplanets undergoing mass and angular momentum accretion. Each protoplanet is modelled as an Accreting sink particle, with $r_{\rm acc} =$ 0.1 \rhill \ (plus signs), 0.5 \rhill \ (diamonds), or 1.0 \rhill \ (asterisks) for each mass. The analytic Type I model of \protect \cite{TanTakWar2002} is plotted as a dashed line. The smallest accretion radii protoplanet models give the migration rates that most closely match the analytic model.}
\label{fig:typeb-radcomp}
\end{figure}

Fig.~\ref{fig:typeb-radcomp} includes all the different accretion radii used in our Accreting sink calculations. It can be seen that those with radii of 0.1\rhill \ (marked as plus symbols) and masses $\lesssim 0.3 \rm \ M_{Jupiter}$ closely follow the analytic model of \cite{TanTakWar2002}. In these cases the dilution of the angular momentum and the effective change in the ratio of $r_{\rm acc}$ to \rhill are minimal due to the small accretion radii, and lower accretion rates. The accretion rates onto these Accreting sinks are still substantial because of the evacuation of gas within $r_{\rm acc}$, which leads to an inward pressure gradient near their boundary where the gas is removed. This effectively pulls material in at a faster rate than would be expected in the absence of the hole. This artificial evacuation at the planet boundary not only leads to artificially fast accretion, it also alters the density structure around the protoplanet out to distances that are large compared to $r_{\rm acc}$; this is tackled in section \ref{sec:isosurfaces}.

\subsection{Self-gravity}

\begin{figure}
\centering
\includegraphics[width=1.0 \columnwidth]{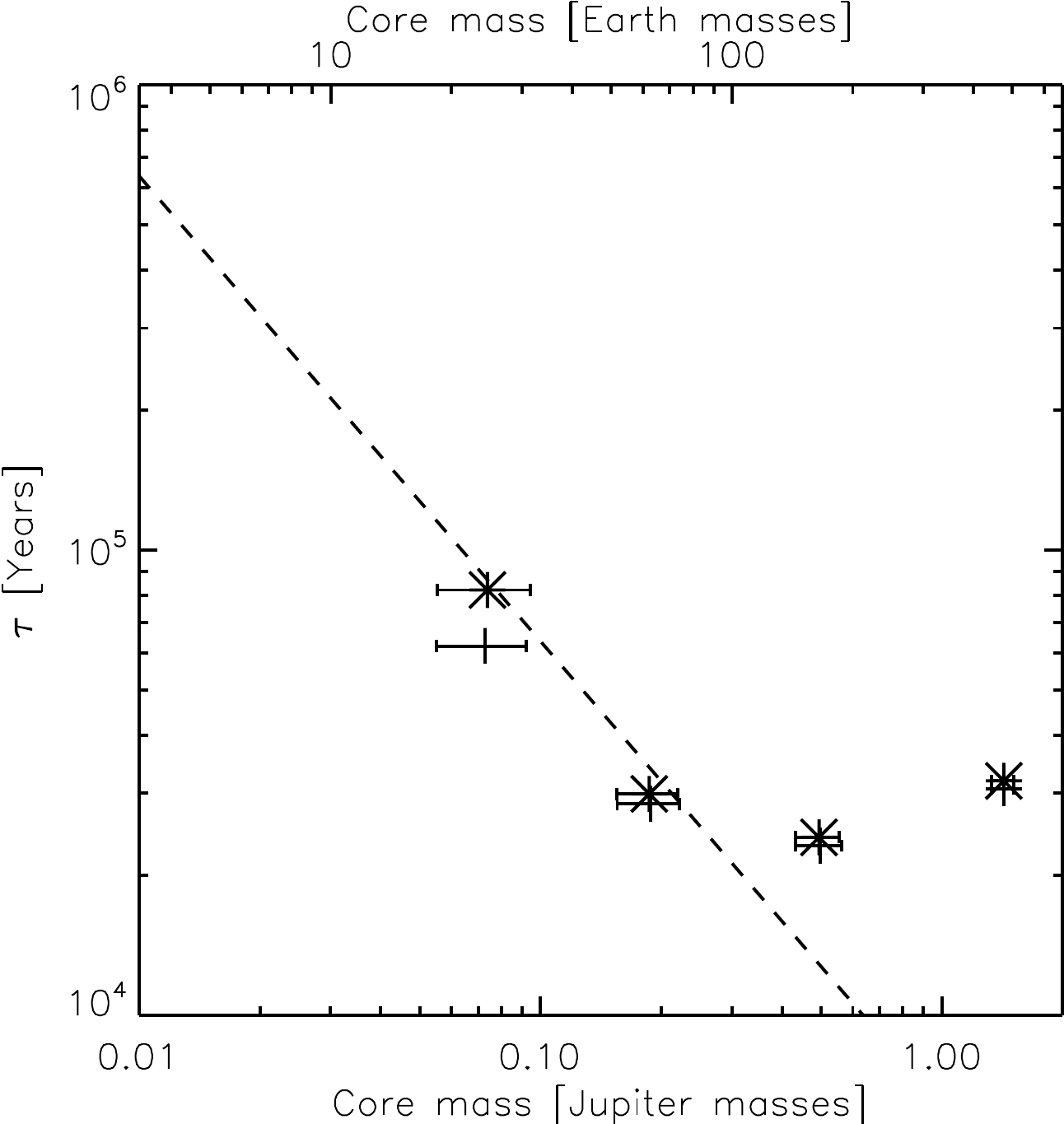}
\caption{Comparisons between the migration timescales of protoplanets, modelled in locally-isothermal discs by Accreting sinks with $r_{\rm acc} = 0.1$ \rhill, in non-self-gravitating discs (plus symbols), and self-gravitating discs (asterisks). As has been found by several authors, the inclusion of disc self-gravity has a very small effect upon the migration timescales. This small impact is consistent across all protoplanet masses, slightly increasing the migration timescales. The analytic Type I model of \protect \cite{TanTakWar2002} is plotted as a dashed line.}
\label{fig:typecvsb}
\end{figure}

All of our models discussed to this point have included the gravitational interaction of the protoplanet with the disc and vice versa, but have neglected the disc's self-gravity, that is its influence upon itself. For a low mass disc this is a reasonable approximation as the central star is so dominant. Indeed the inclusion of self-gravity in low-mass discs has been found to have only a small impact on migration rates by several authors \citep{NelBen2003, PieHur2005, BarMas2008, CriBarKleMas2009}. Comparing two series of identical calculations, with the inclusion of self-gravity being the only differentiating characteristic, we found there to be a very small, but consistent increase in the migration timescales for self-gravity models. Fig.~\ref{fig:typecvsb} makes evident the small magnitude of the change due to self-gravity, in this instance for Accreting sinks with $r_{\rm acc} = 0.1$ \rhill. However, self-gravity remains an essential component in building a self-consistent model. 

\begin{figure}
\centering
\includegraphics[width=1.0 \columnwidth]{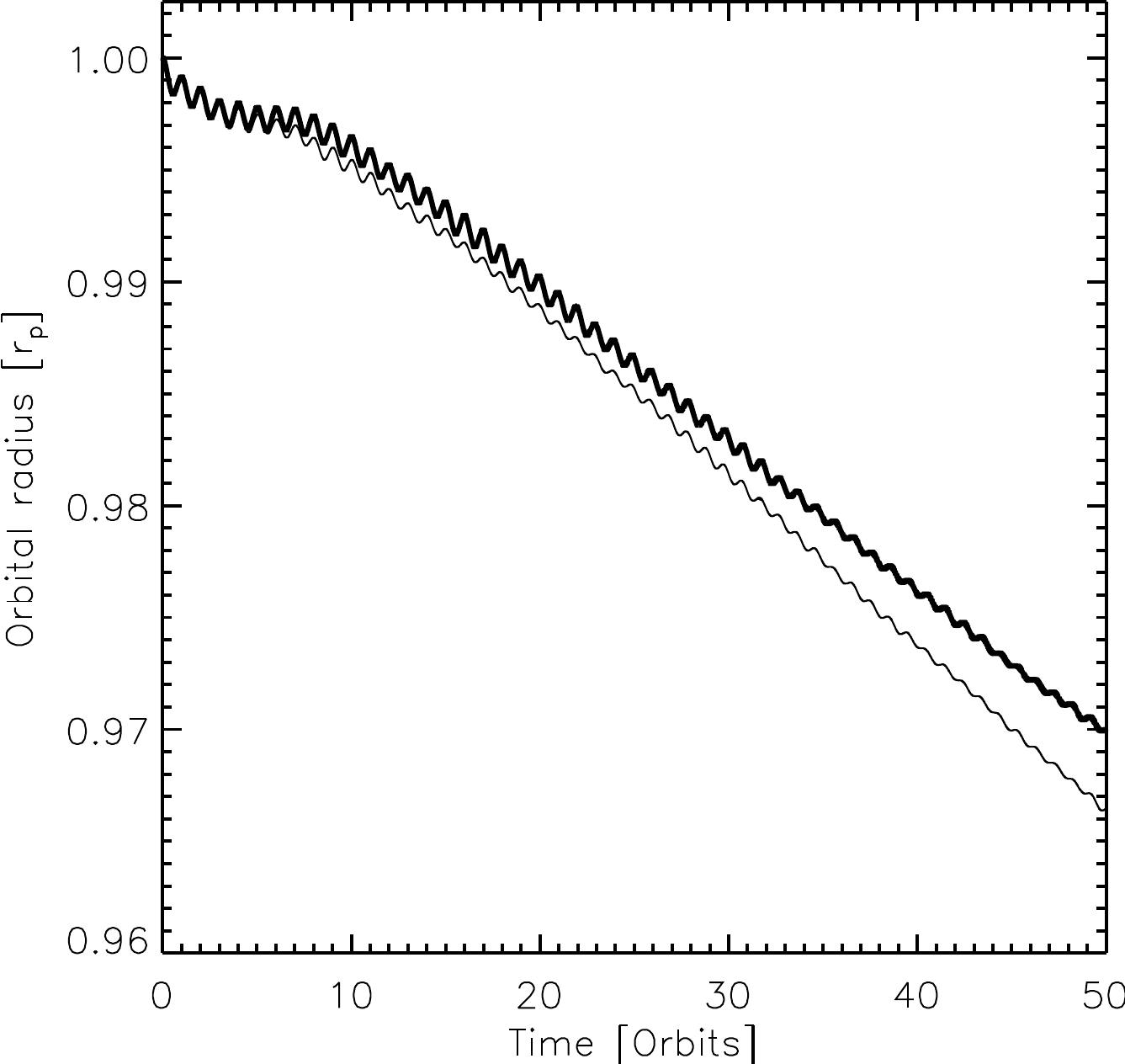}
\caption{The orbital evolution of a 333 \earthmass \ protoplanet modelled using a point mass with surface of radius 0.03 \rhill, using a locally-isothermal equation of state. A calculation including disc self-gravity is shown using a narrow line, whilst the non-self-gravity calculation is shown using a thick line. There is reduction in the migration rate of the protoplanet modelled without self-gravity as it develops a circumplanetary disc; a result of the inertial mass problem. The migration rate is slower by $\approx$ 14 per cent.}
\label{fig:inertial}
\end{figure}

When substantial circumplanetary discs or envelopes form, the inclusion of self-gravity becomes important in delivering accurate migration rates. As discussed in \cite{CriBarKleMas2009}, a circumplanetary disc is locked to the protoplanet, and moves with it around the central star. In the absence of the disc self-gravity the circumplanetary disc does not experience the torques due to the protoplanetary disc; the torques which the protoplanet is responding to. As such it does not migrate of its own accord, but must be pulled by the protoplanet, artificially slowing the rate of migration; known as the `inertial mass problem' \citep{CriBarKleMas2009}. To explore the impact of this phenomenon we performed two calculations using point masses with surfaces (discussed in the next section) to model 333 \earthmass \ protoplanets, one model with and one without the inclusion of disc self-gravity. The two resulting migration evolutions are shown in Fig.~\ref{fig:inertial}. In both cases a protoplanetary disc develops, and in the case without disc self-gravity, this causes the protoplanet's rate of migration to slow by approximately 14 per cent. Having illustrated the effect, we proceed with calculations which all include self-gravity to avoid what is an unphysical interaction.

\begin{figure}
\centering
\includegraphics[width=1.0 \columnwidth]{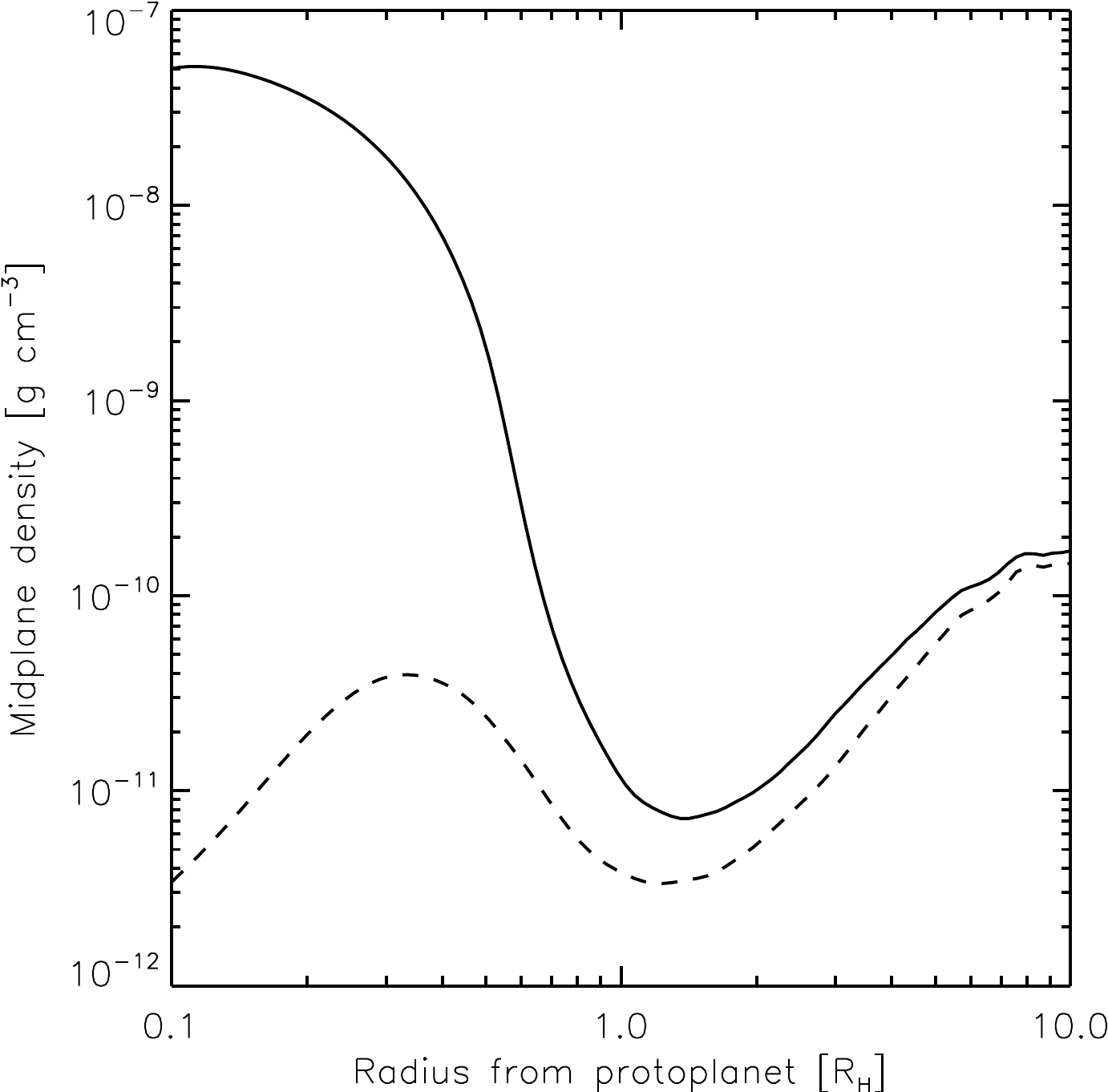}
\caption{Midplane densities around a 333 \earthmass \ protoplanet modelled using a locally-isothermal equation of state and self-gravity after 50 orbits. The solid line denotes the densities around a protoplanet modelled using a point mass with a surface of radius 0.1 \rhill. The dashed line gives the midplane densities around an identical protoplanet modelled by an Accreting sink with $r_{\rm acc} = 0.1$ \rhill. The evacuation of material at the Accreting sink's boundary leads to an inward pressure gradient that pushes in gas, giving artificially high accretion rates, and reducing the density in the vicinity of the growing planet out to radii of $\gtrsim $ \rhill.}
\label{fig:sinksurf}
\end{figure}

\begin{figure*}
\centering
\includegraphics[width=1.0 \textwidth]{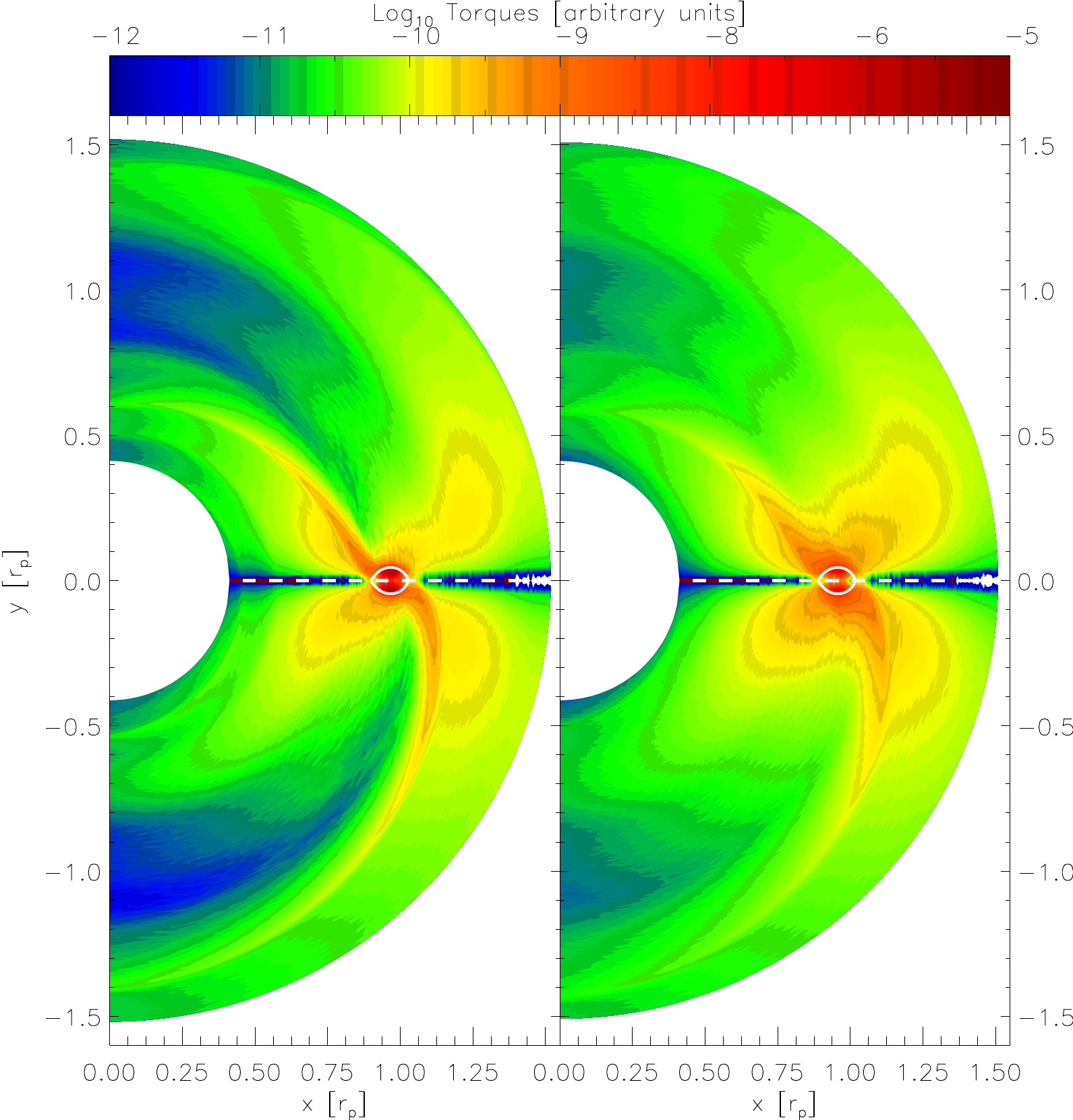}
\caption{Torque distributions around a 333 \earthmass \ protoplanet modelled by a point mass with a surface of radius 0.03 \rhill \ after 50 orbits. The left hand panel is from a locally-isothermal model, whilst the right hand panel is from a radiation hydrodynamic calculation using interstellar opacities; both cases include self-gravity, and the figures exclude torques from within \rhill/3. The Roche lobe is marked on in each case using a white line. All the torques above the white dashed line are positive, whilst all those below are negative. Introducing radiative transfer smears out the spiral arms, fills in the corotation region, and reduces the maximum torques which act from within the Roche lobe. The section shown is $r_{\rm p} \pm 8$ \rhill, which avoids the inner and outer boundary regions.}
\label{fig:torquemap}
\end{figure*}

\begin{figure*}
\centering

\subfigure 
{
\setlength\fboxsep{0pt}
\setlength\fboxrule{0.0pt}
\fbox{\includegraphics[height=18cm]{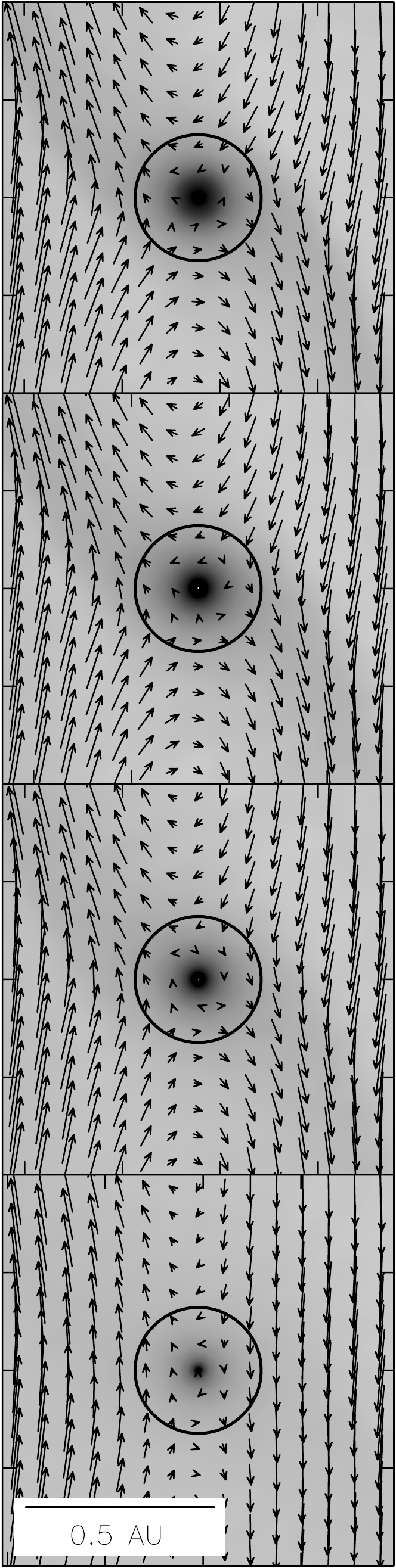}}
}
\nolinebreak
\hspace{-4.5mm}
\subfigure 
{
\setlength\fboxsep{0pt}
\setlength\fboxrule{0.0pt}
\fbox{\includegraphics[height=18cm]{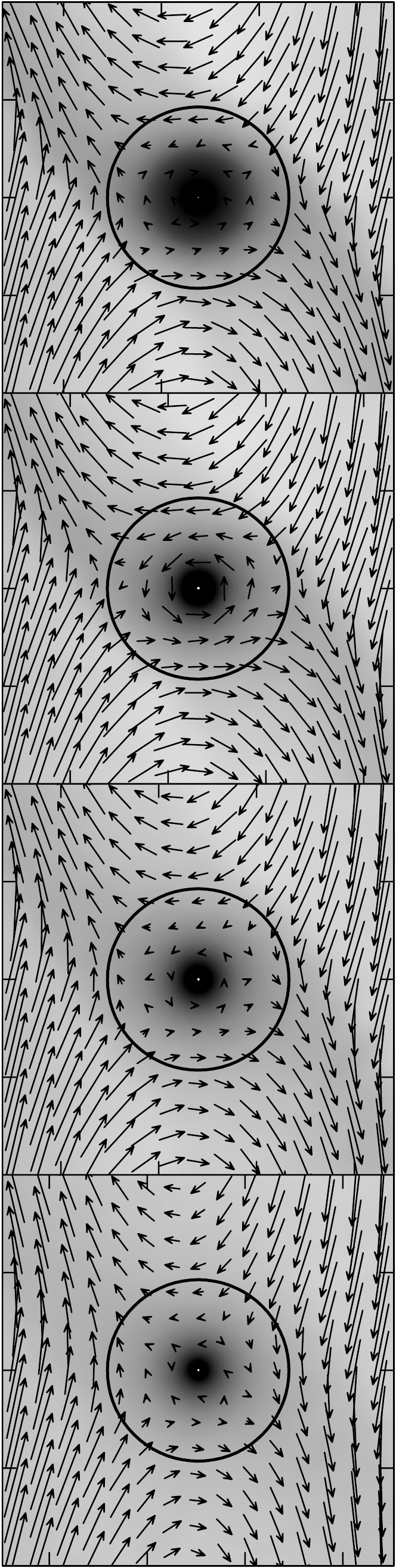}}
}
\nolinebreak
\hspace{-4.5mm}
\subfigure 
{
\setlength\fboxsep{0pt}
\setlength\fboxrule{0.0pt}
\fbox{\includegraphics[height=18cm]{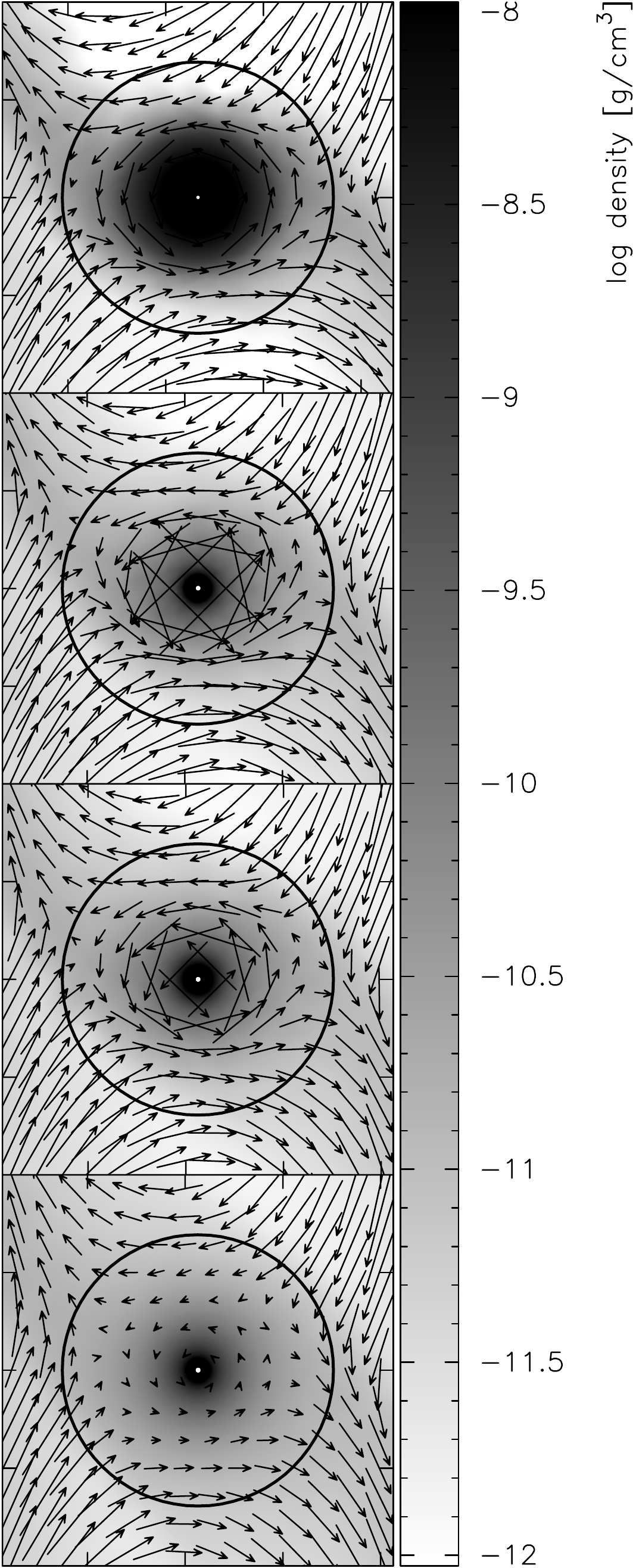}}
}

\caption{Sections of the whole disc models plotted as midplane density maps, focussing on the region surrounding the embedded protoplanets, with velocity vectors overplotted. It can be seen that material is flowing into the Hill sphere and back out again, reaching significant depths within this region. This material is able to exchange angular momentum with the protoplanet before escaping its potential, and so a persistent torque is established from within \rhill. From left to right the protoplanet masses are 33, 100, and 333 \earthmass. From top to bottom the models are locally-isothermal, and then radiation hydrodynamical with 1, 10, and 100 per cent interstellar grain opacities.}
\label{fig:denszoom}
\end{figure*}

\subsection{Protoplanets with surfaces}
\label{sec:isosurfaces}

The final addition to our locally-isothermal models was to use our planet surface treatment. This addresses the failing of our Accreting sink models by allowing the gas in the protoplanet's vicinity to develop a structure free of the undesired effects of evacuating gas near the protoplanet. Combining this treatment with the inclusion of self-gravity we obtain a model in which the interaction of the disc and protoplanet is free of any artificially imposed radial limit, such as excluding material within \rhill. Instead, the radius at which gas becomes bound to the protoplanet, and stops applying orbit affecting torques is achieved self-consistently. Fig.~\ref{fig:sinksurf} illustrates the midplane density distribution around both an Accreting sink and a point mass with surface (in both cases a 333 \earthmass \ protoplanet, of radius 0.1 \rhill). Whilst gas locked in a circumplanetary disc will not exert migration causing torques, gas beyond such a disc (found to be $\gtrsim $\rhill/3 in these models, \citealt{AylBat2009b}) that moves in and out of the Hill sphere will. The density distribution in the two cases are not similar until radii of $\gtrsim$ \rhill, suggesting that the Accreting sinks experience smaller torques from within and around \rhill \ than the point masses with surfaces. The importance of these local torques can be seen in the left panel of Fig.~\ref{fig:torquemap} which is a map of the torques acting on the protoplanet in the plane of the disc. The figure excludes torques acting from within \rhill/3 which is the expected limit of the circumplanetary disc for such a high mass protoplanet, but note that within the model only the radius of the planet's surface is specified. Between \rhill/3 and the Roche lobe (outlined in white) are the strongest interactions by virtue of the gas's proximity and density. These torques are enabled by gas flow into and out of the Roche lobe which allows for the continuous exchange of angular momentum, establishing a persistent torque. This flow can be seen for locally-isothermal models in the top panels of Fig. \ref{fig:denszoom}. The velocity vectors are plotted over a midplane density map of the area surrounding the protoplanet. Gas can be seen passing into and out of the Hill sphere on trajectories that are both radial (horseshoe orbits) and azimuthal.

\begin{figure}
\centering
\includegraphics[width=1.0 \columnwidth]{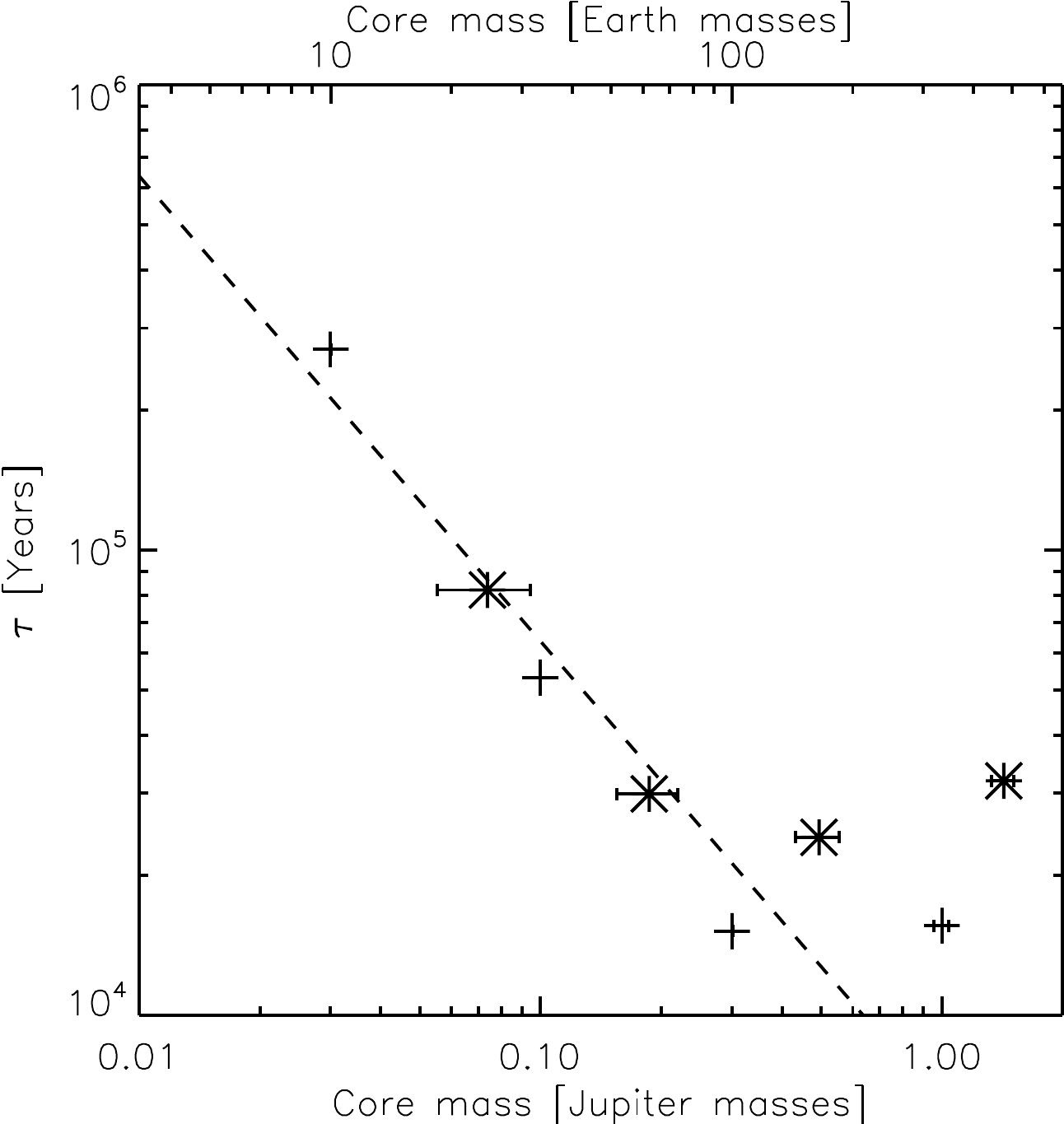}
\caption{The migration timescales of protoplanets modelled using Accreting sinks (asterisks) compared with those obtained using point masses with surfaces (plus signs), with accretion/surface radii of 0.1 \rhill. These timescales are from models of locally-isothermal self-gravitating discs. The analytic Type I model of \protect \cite{TanTakWar2002} is plotted as a dashed line. With small accretion radii, the Accreting sinks give results inline with the analytic expectation for Type I migration and with the more realistic point mass with surface models. Beyond a Saturn mass, in the Type II regime, the Accreting sinks show slower migration by factors of $\approx 2$.}
\label{fig:typedvsc}
\end{figure}

Fig.~\ref{fig:typedvsc} includes migration timescales from Accreting sink models and point mass with surface models with accretion/surface radii of 0.1 \rhill. Throughout the low mass regime both types of model lead to results in good agreement with the analytic model of Type I migration. At and beyond a Saturn mass, where Type II migration becomes effective, the point mass with surface calculations show slightly faster ($\approx$ factor of 2) rates of migration. This is a result of the less evacuated gaps formed in sink with surface models, which results in a faster migration mechanism which is intermediate between Type I and II. The gap is more thoroughly evacuated when using an Accreting sink because it establishes no envelope structure to hinder accretion, and a negative pressure gradient to encourage it. This strips the corotational region of material quickly, resulting in a clearer gap at an equivalent time. However, in both cases the gap is becoming more evacuated with time, which would be expected to slow migration to a Type II rate; this is a result of the models relatively short evolution of just 50 orbital periods starting from a uniform disc.


\subsection{Protoplanets with surfaces and radiation hydrodynamics}

The structure of protoplanetary discs, and circumplanetary envelopes and discs, are the result of competing gravitational and thermal effects. In the locally-isothermal models discussed to this point, the thermal effects associated with embedding a protoplanet in a protoplanetary disc have not been accounted for. To address this shortcoming we finally introduce radiative transfer and a realistic equation of state into the migration models. The inclusion of a better thermodynamic treatment has several potentially important ramifications. The accretion energy released near a growing protoplanet warms its surroundings, which in concert with disc self-gravity leads to the development of self-consistent local disc and envelope structures. For example, a more poorly defined disc gap (opened by high mass planets) develops when radiative transfer is included \citep{FouMay2008,AylBat2009}. Compressional heating also tends to smear out the spiral density waves that are launched by planets embedded in the discs, giving them lower peak densities than in locally-isothermal models. 

\begin{figure*}
\centering
\subfigure 
{
    \includegraphics[width=1.0 \columnwidth]{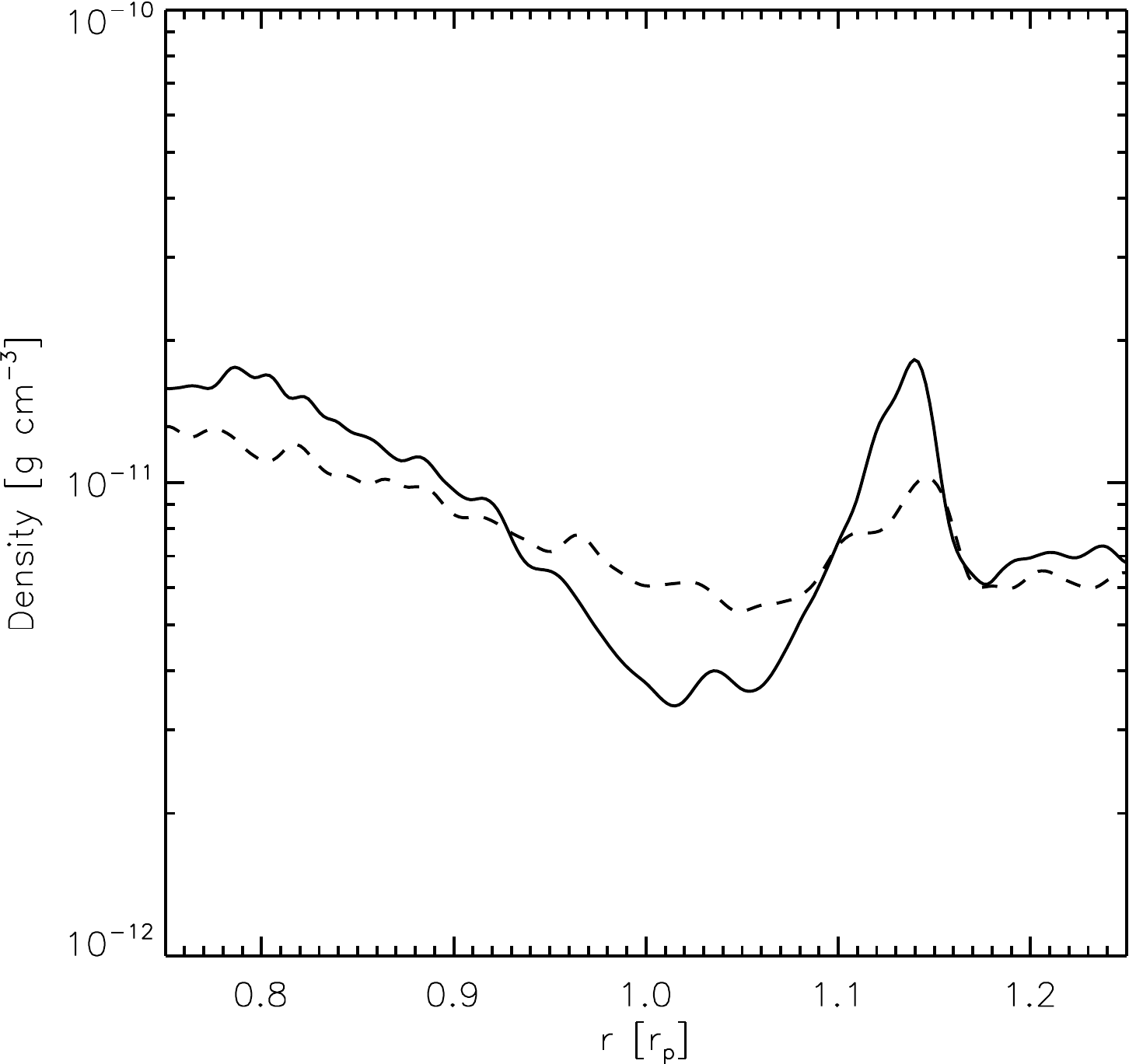}
}
\subfigure 
{
    \includegraphics[width=1.0 \columnwidth]{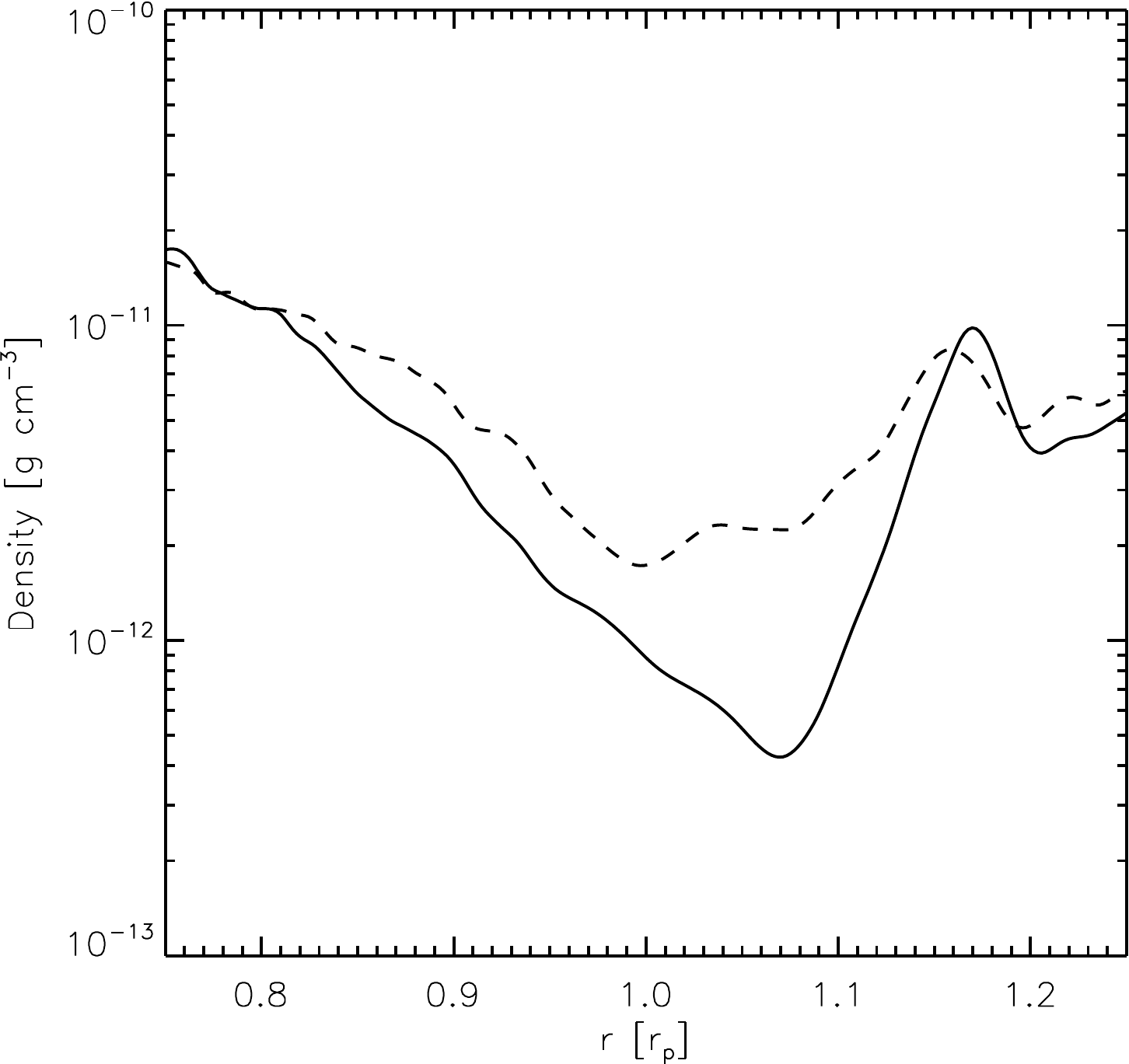}
}
\caption{Midplane densities along lines 10 degrees clockwise of 100 \earthmass \ (left) and 333 \earthmass \ (right) protoplanets, modelled with point masses with surfaces of 0.03 \rhill \ after 50 orbits. The profiles run over $\rm r_{p} \pm 0.25 r_{p}$. The solid lines mark the profiles for locally-isothermal calculations, whilst the dashed lines are taken from radiation hydrodynamics calculations using interstellar grain opacities. In both cases, the introduction of radiative transfer leads to poorer evacuation of the corotation region, allowing for greater torques to act from these radii. For the 100 \earthmass \ protoplanet the outer spiral arm can be seen to have a lower peak density, a result of the heat produced in the shock.}
\label{fig:isortline}
\end{figure*}

The impact of these changes can be seen in the panels of Fig.~\ref{fig:torquemap}, which illustrate the torques acting on a 333 \earthmass \ protoplanet in both a locally-isothermal model (left panel) and an interstellar opacity radiation hydrodynamical (RHD) model (right panel). The peak torques acting near the protoplanet are lower in the RHD case, though the maxima occur in similar locations (near the Roche lobe), illustrating the impact of the reduced local density. In the locally-isothermal case, the azimuthal angle at which torques from the corotation region fall away is considerably smaller than in the RHD case. This is a result of the more poorly cleared gap when using RHD. Finally, the thermal spreading of the spiral arms can be seen by the broadened regions from which the associated torques are exerted, though there is no discernible difference in the magnitude of the torque acting from wave crest. Densities along a line in the midplane, 10 degrees clockwise of the same 333 \earthmass \ protoplanet in both locally-isothermal, and RHD models, are shown in the right hand panel of Fig.~\ref{fig:isortline}. This distribution directly illustrates the difference in the vacuity of the gap. It is also clear that the density maximum of the (outer) spiral arm is similar in both instances, despite the breadth being altered by the heating. The left panel is an equivalent plot for a 100 \earthmass \ protoplanet. In this case the spiral arm density maximum is reduced by about a factor of 2 compared with the locally-isothermal case, possibly weakening the differential Lindblad torques. However, the disc gap contains more mass than in the locally-isothermal model, once again leading to a migration process even more similar to Type I, and so faster, than the expected Type II. The overall impact of using radiation hydrodynamics in this case will depend on the relative impact of these two opposing effects.

\begin{figure}
\centering
\includegraphics[width=1.0 \columnwidth]{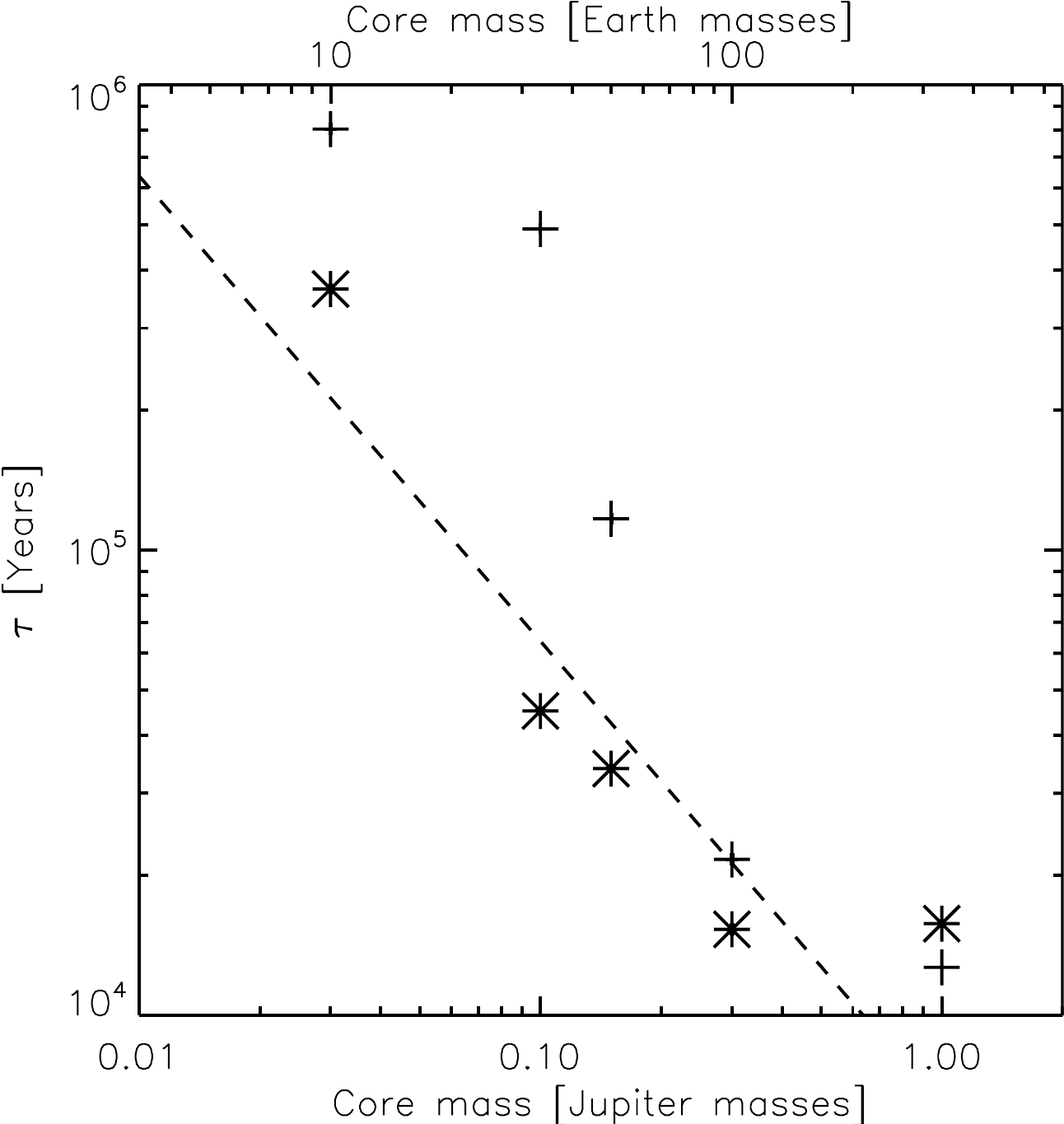}
\caption{The migration timescales of protoplanets modelled using point masses with surfaces of radius 0.03 \rhill \ in locally-isothermal (asterisks) and interstellar grain opacity radiation hydrodynamics (plus symbols) models. The analytic Type I model of \protect \cite{TanTakWar2002} is plotted as a dashed line. In the mass regime corresponding to Type I migration, the inclusion of radiative transfer with high opacities leads to slower rates of migration. For a 33 \earthmass \ protoplanet the migration timescale is increased by an order of magnitude. Such slowed migration may improve the chances of proto-giant cores surviving long enough to grow to giant planet masses.}
\label{fig:typeevsd}
\end{figure}

\begin{figure}
\centering
\includegraphics[width=1.0 \columnwidth]{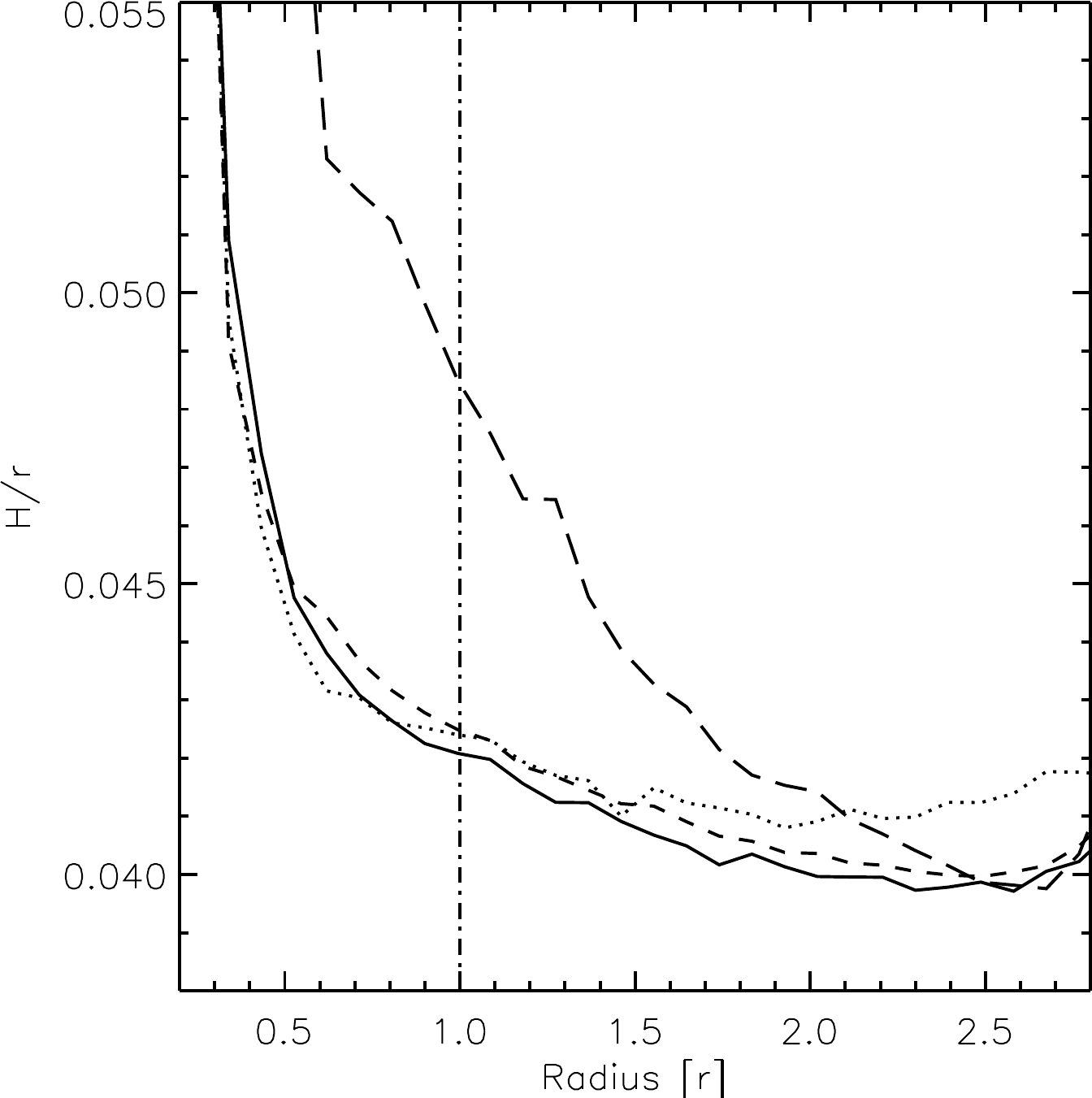}
\caption{Scaleheights measured in discs containing a 33 \earthmass \ protoplanet modelled using a point mass with surface of radius 0.03 \rhill. The results of a locally-isothermal model are shown using a solid line, along with the radiation hydrodynamics models using 100 per cent (long-dashed), 10 per cent (short-dashed), and 1 per cent (dotted) interstellar grain opacities. The locally-isothermal and reduced opacity radiation hydrodynamics models all yield similar scaleheights through the disc, but the full opacity model gives a thicker disc, most notably around the radius of the embedded protoplanet.}
\label{fig:scaleheights}
\end{figure}

The migration rates found in these interstellar grain opacity radiation hydrodynamical models are compared with those from otherwise equivalent locally-isothermal models in Fig.~\ref{fig:typeevsd}. In the high mass regime, the impact is very small, with migration rates altered by no more than 30\%. The 333 \earthmass \ protoplanet, which maintains the same spiral arm densities but a less evacuated gap, migrates faster with RHD. This is consistent with an intermediate migration process, between Types I and II, as a result of the less evacuated gap formed with RHD. Once again, it is important to note that for both high mass protoplanets the change in migration timescales when radiative transfer is introduced are very small. In the lower mass Type I regime, the introduction of radiative transfer leads to substantial increases in the migration timescales. A 50 \earthmass protoplanet takes almost 4 times longer to migrate into its central star when modelled with radiative transfer than it does without, whilst a  33 \earthmass \ protoplanet takes more than 10 times longer. A 10 \earthmass \ protoplanet is slowed by more than a factor of 2. In these cases, the densities around the protoplanets out to several \rhill \ are lower when modelled with RHD using interstellar grain opacities than in the locally-isothermal models; about a 33 \earthmass \ protoplanet the gas density at the disc midplane at a radius of \rhill \ is a factor of 3 lower. 
Similar reductions in Type I migration rates have been found in other recent work considering non-isothermal discs \citep[e.g.][]{PaaMel2006}. Recent models suggest that the slowing results from the dependence of the co-orbital torques on the radial entropy gradient of the protoplanetary disc, and that under some circumstances these torques can be sufficient to reverse the direction of migration \citep{PaaPap2008, BarMas2008, KleCri2008, KleBitKla2009, PaaBarCriKle2010}. The reduced migration rates for low mass protoplanets are an important result in trying to increase the survival probability of proto-giant cores. Specifically it may justify the reduction, or some fraction of the reduction in Type I migration rates assumed in core accretion models that seek to ensure planets survive \citep{AliMorBen2004, AliMorBenWin2005, IdaLin2008, MorAliBen2009}.

In the point mass with surface models the envelope/disc surrounding the protoplanets is structured realistically due to the influences of gravity, self-gravity, and thermodynamics. It is therefore possible to account for the differentiation of the angular momentum of accreted material into the planet's orbital angular momentum, and into rotation about the planet. This differentiation has next to no impact upon the compound protoplanet's orbital angular momentum. Typically a fraction of $\sim 10^{-4}$ of the angular momentum of the gas accreted into a circumplanetary disc is diverted into the rotation of that disc about the planet.


The impact of radiative transfer may go beyond the immediate vicinity of a protoplanet, for example, inflating a protoplanetary disc's scaleheight as a result of viscous heating at the midplane. Thicker discs result in smaller torques from the inner and outer Lindblad resonances, scaling as $(H/R)^{-3}$ \citep{Ward1997}, whilst the mismatch between the torques scales as $H/R$. This means that the migration rates scale as $(H/R)^{-2}$, and a hotter, thicker disc should lead to slower migration \citep{MasKle2006}. Fig.~\ref{fig:scaleheights} shows the measured value of $H/R$ against radius for a disc with an embedded 33 \earthmass \ protoplanet. It can be seen that there is a larger scaleheight at and about the protoplanet's orbital radius in a radiation hydrodynamics calculation using interstellar opacities when compared with all the other calculations. The scaleheight is a factor of $\approx 1.2$ greater, which can only account for an increase in the migration timescale of about 50 per cent.

\subsubsection{The impact of opacity in RHD models}
\label{sec:radiative}

\begin{figure}
\centering
\includegraphics[width=1.0 \columnwidth]{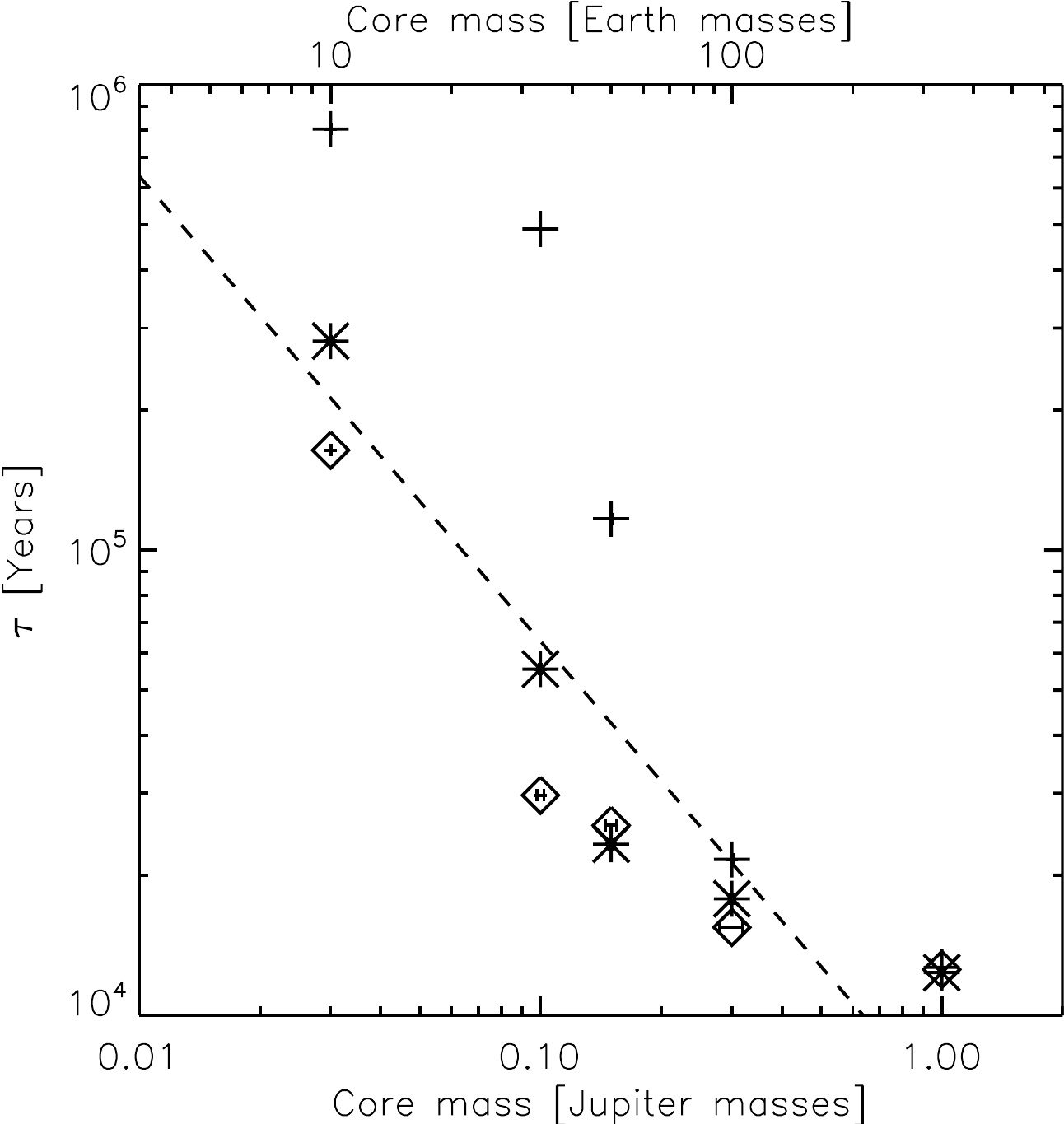}
\caption{Migration timescales for protoplanets in radiation hydrodynamics calculations, modelled by point masses with surface radii of 0.03 \rhill, with three different opacities: 100 per cent (plus symbols), 10 per cent (asterisks), and 1 per cent (diamonds) interstellar grain opacities. The reduced opacity models yield shorter migration timescales than their locally-isothermal equivalent models (c.f. Fig.~\ref{fig:typeevsd}), with only the full opacity case leading to slower migration. Increasing the grain opacity of the disc leads to slower migration rates, with the effect becoming smaller for higher mass protoplanets. The analytic Type I model of \protect \cite{TanTakWar2002} is plotted as a dashed line.}
\label{fig:typeevsfvsg}
\end{figure}

Our final results come from varying the protoplanetary disc's opacity to determine its effect upon migration. Fig.~\ref{fig:typeevsfvsg} shows that the effect of varying the opacity on the migration rates is very small for the high mass planets. For the low mass planets there is a trend of increasing migration timescale with increased opacity. Interestingly, protoplanets in the low opacity RHD models migrate faster than in the locally-isothermal models.

\begin{figure*}
\centering
\includegraphics[width=1.0 \textwidth]{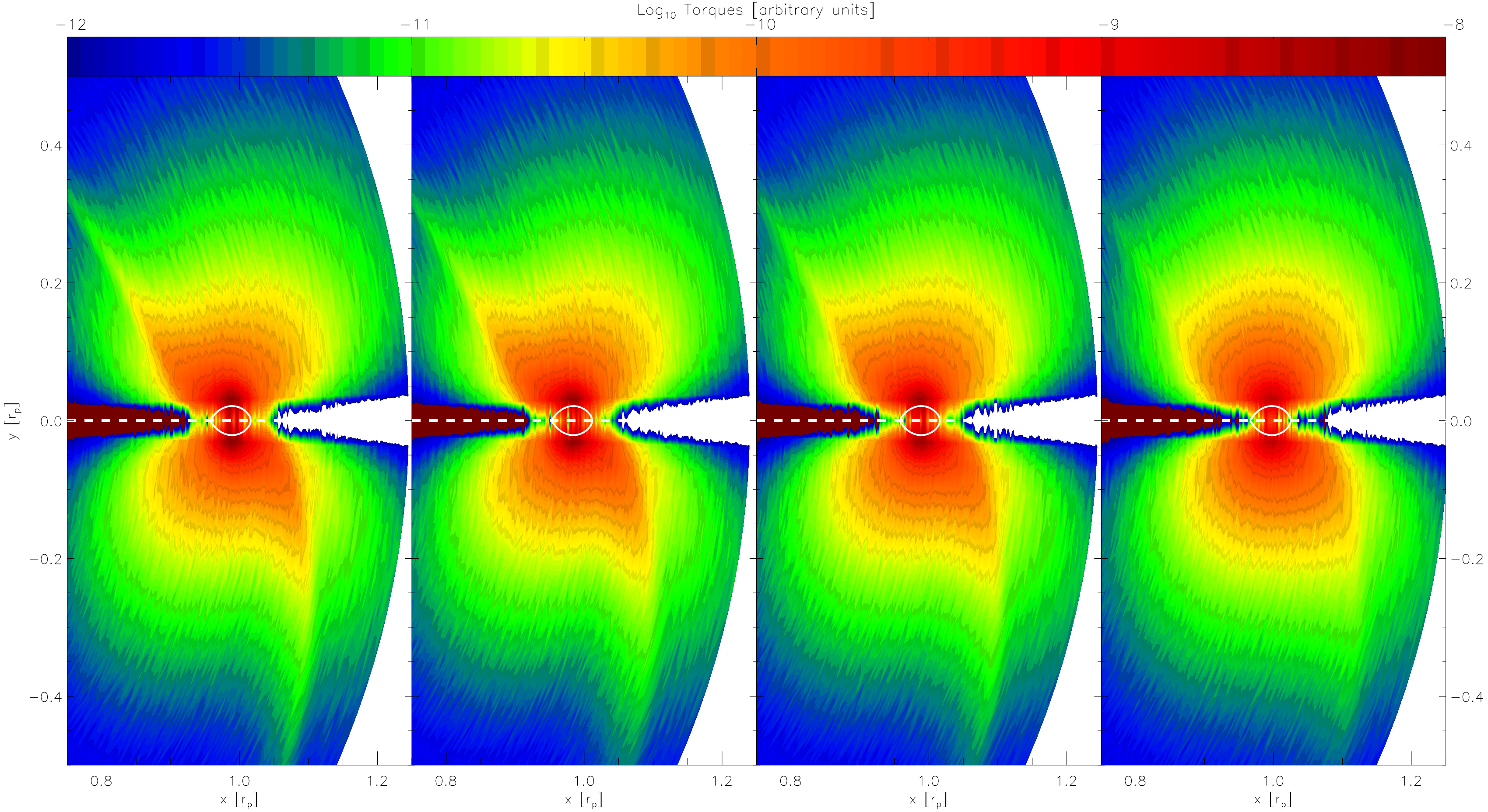}
\caption{Torque distributions around a 33 \earthmass \ protoplanet modelled by a point mass with a surface of radius 0.03 \rhill \ after 50 orbits. From left to right the models are locally-isothermal, and then RHD with 1 per cent, 10 per cent, and 100 per cent interstellar grain opacity respectively, all including self-gravity. The Roche lobe is marked on in each case using a white line. All the torques above the white dashed line ($y = 0$) are positive, whilst all those below are negative. The spiral arms are particularly poorly defined in the interstellar grain opacity case (right most panel), and the torques acting from around the Roche lobe are weakest in this case.}
\label{fig:torquecomp}
\end{figure*}

The largest change in the migration timescales of low mass protoplanets is seen when employing the interstellar grain opacities, with the change between the 1\% and 10\% models, and the locally-isothermal models being somewhat smaller. In all but the interstellar grain opacity calculations, the densities around a given protoplanet are almost identical, and so the local torques should be similar. The spiral arms are also likely to be most broadened, and their peak density most reduced, in the highest opacity calculations, with lower opacities leading to structures more similar to those found in the locally-isothermal models. These structures, and maximum local torques can be seen in the torque maps shown in Fig.~\ref{fig:torquecomp}, which are produced for a 33 \earthmass \ protoplanet modelled using a point mass with a surface radius of 0.03 \rhill. The torques acting at and around the Roche lobe can be seen to be weaker in the interstellar opacity case (right most panel), than in any of the other cases, and the spiral arms most diminished.

\begin{figure*}
\centering
\includegraphics[width=0.85 \textwidth]{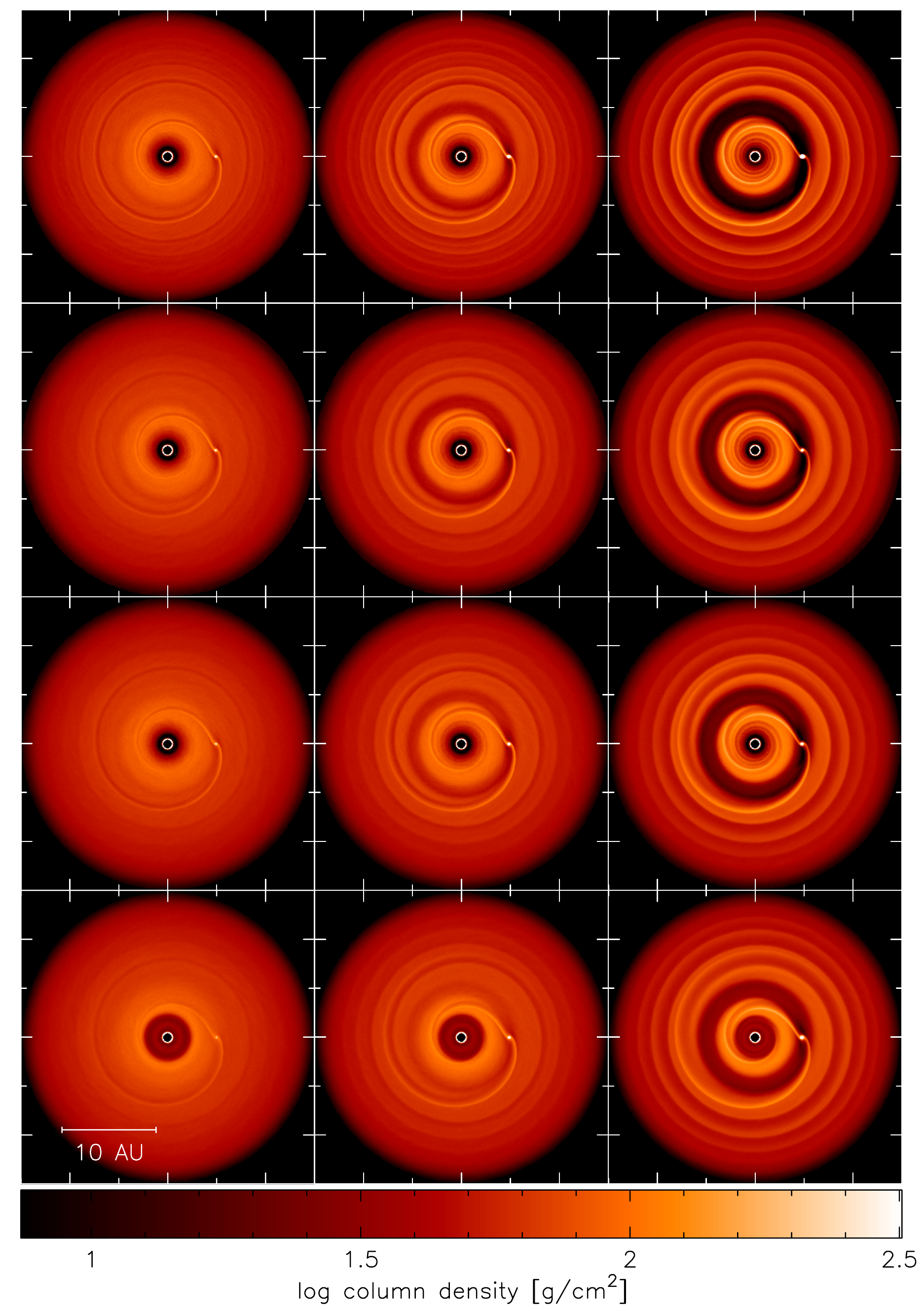}
\caption{Surface density maps of protoplanetary discs after 50 orbits of the embedded protoplanets. From left to right the protoplanet masses are 33, 100, and 333 \earthmass. From top to bottom the models are locally-isothermal, and then radiation hydrodynamical with 1, 10, and 100 per cent interstellar grain opacities.}
\label{fig:pM333dens}
\end{figure*}

\begin{figure*}
\centering
\includegraphics[width=0.85 \textwidth]{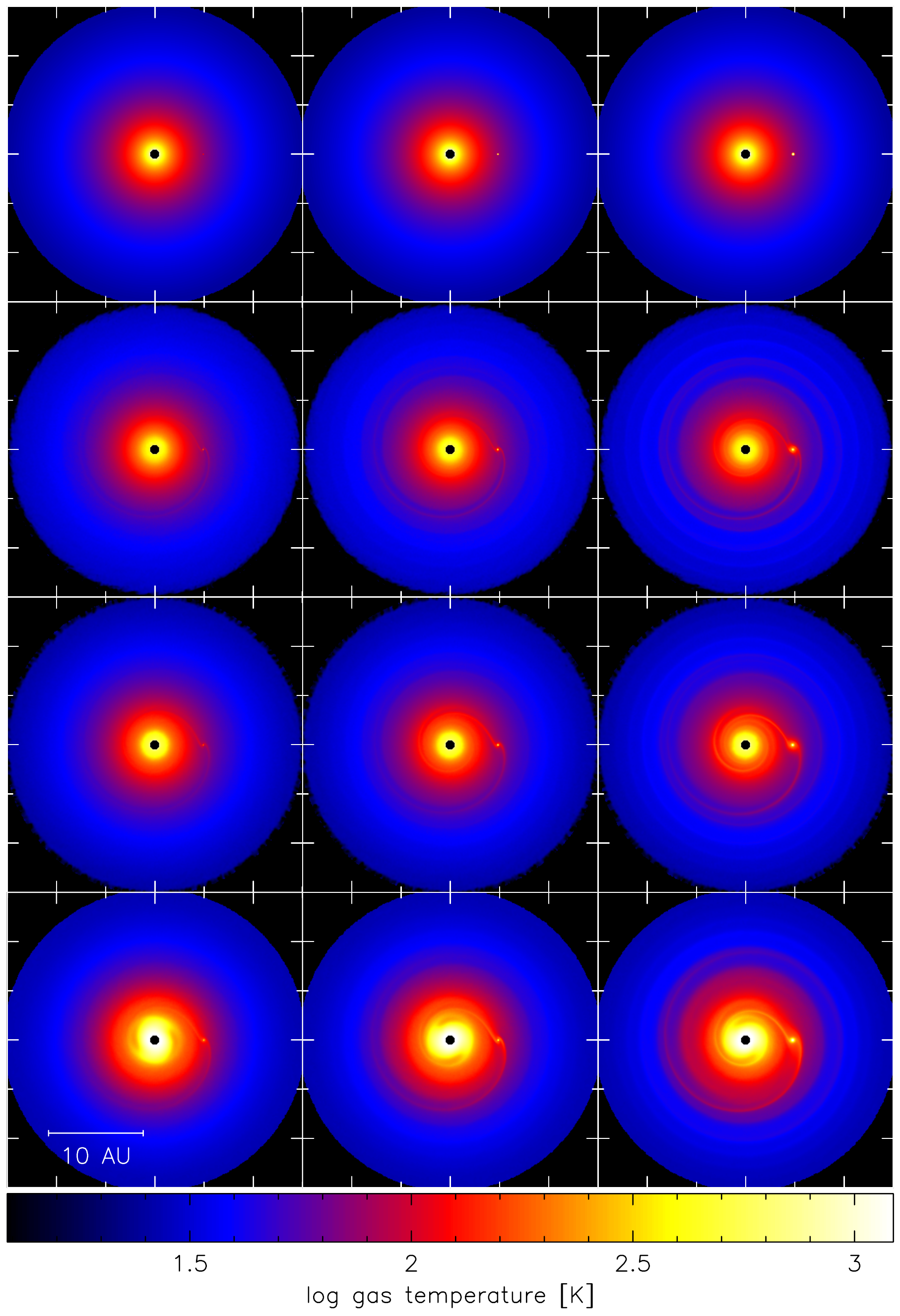}
\caption{Density weighted temperature maps of protoplanetary discs after 50 orbits of the embedded protoplanets. From left to right the protoplanet masses are 33, 100, and 333 \earthmass. From top to bottom the models are locally-isothermal, and then radiation hydrodynamical with 1, 10, and 100 per cent interstellar grain opacities.}
\label{fig:pM333temp}
\end{figure*}

\begin{figure*}
\centering
\includegraphics[width=0.85 \textwidth]{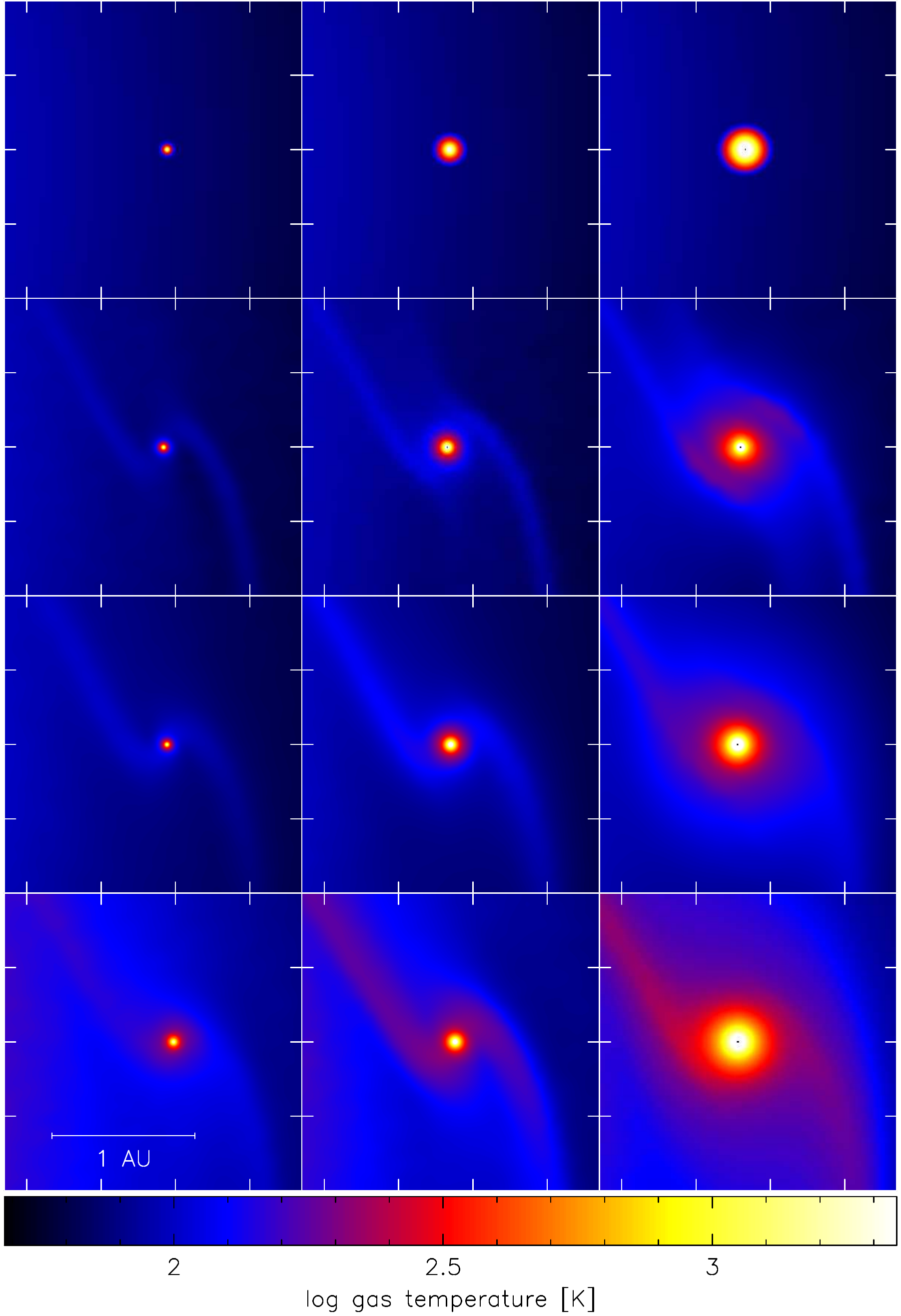}
\caption{Magnified sections of the whole disc midplane temperature maps shown in Fig.~\ref{fig:pM333temp}, focussing on the region surrounding the embedded protoplanets. Again, from left to right the protoplanet masses are 33, 100, and 333 \earthmass. From top to bottom the models are locally-isothermal, and then radiation hydrodynamical with 1, 10, and 100 per cent interstellar grain opacities. It is particularly apparent for the 333 \earthmass \ protoplanet that the energy released as a result of accretion leads to a larger region of hot gas, which is less able to cool as the opacity is increased. It is also possible to see the broadening of the spiral arms as the opacity is increased.}
\label{fig:tempzoom}
\end{figure*}

The diminution of the spiral arms with increasing opacity can be further seen in Figs.~\ref{fig:pM333dens} and \ref{fig:pM333temp}, which are surface density and density weighted temperature maps respectively of discs with embedded protoplanets, once again modelled by point masses with surfaces of radius 0.03 \rhill. From left to right the protoplanet masses are 33, 100, and 333 \earthmass  \ respectively, and from top to bottom the models are locally-isothermal, and RHD with 1, 10, and 100 per cent interstellar grain opacities respectively; we have omitted the 10 \earthmass \ cases because the disturbance caused by the protoplanet is barely visible. Moving from the locally-isothermal density distribution (top-row of figure \ref{fig:pM333dens}) to the interstellar opacity case (bottom row) it is evident that where a gap forms, it is less clear at higher opacities, and that the density waves propagating outwards are less peaked. Fig.~\ref{fig:denszoom} shows midplane densities and velocity vectors in the vicinity of the protoplanets. The velocity vectors make plain the spiral shocks, horseshoe orbits, and the transition to circumplanetary flow deep within the Hill sphere. \ref{fig:tempzoom} shows density weighted temperature maps for the same models  but concentrating on just the area surrounding the protoplanet.

The diminution of the propagating density waves with increasing opacity maybe in part due to the process of wave channelling. In a vertically isothermal disc, these density perturbations propagate as plane waves \citep{LubPri1993}. However, once radiative transfer is introduced, a vertical temperature profile is established, in which the midplane of the disc is hottest due to the effects of compression and viscous heating. This leads to wave channelling, causing the energy of the wave to quickly become confined to the surfaces of the disc, where it is dissipated \citep{KorPri1995, LubOgi1998, OgiLub1999, BatOgiLubPri2002}. This results in the more limited propagation of the spiral arms seen in the non-isothermal models, and which is clearest for low mass planets where the weak spirals barely last an entire planetary orbital period. It is also possible to see in Fig.~\ref{fig:denszoom} that the breadth of the spiral arms increases with opacity, as was discussed earlier.

In Fig.~\ref{fig:pM333temp} and Fig.~\ref{fig:tempzoom} it is possible to see the hot gas that develops around the protoplanets, particularly those of higher mass, when modelled with RHD and interstellar grain opacities. In Fig.~\ref{fig:pM333temp} it is also possible to see the heat produced in the spiral shocks. The spatial extent of the hot gas surrounding a protoplanet reduces with the reduction of the opacity, which allows the less dense outer reaches to cool more easily. It is also possible to see the higher temperatures in the spiral shocks, even in those waves that have propagated far from the planet, when compared with the underlying temperature distribution shown in the top row.

\section{Discussion}
\label{sec:migdiscussion}

It seems that when planets form, they should also migrate. This migration poses problems for the survival of gas giant cores. Conversely, such migration may aid in planet growth by opening up undepleted regions of the disc for these protoplanets to accrete \citep{AliMorBen2004}. It is known from the ever increasing number of exoplanet detections that a large population of planets do grow and survive around many different types of star. The work discussed in this paper, like the work of many authors, was undertaken to try and ascertain the influence of different physical processes upon migration. The models we present in section \ref{sec:radiative} are the most self-consistent to date, including self-gravity, radiative transfer, and protoplanets modelled with surfaces to produce realistic envelopes and circumplanetary discs.

\subsection{Code comparison}
We first made comparisons with the results of a grid code to ensure that our SPH model was suited to tackling the problem of migration, a capacity in which it has been seldom used before.  The ability of the SPH method to model protoplanet migration was considered by \cite{deVEdgArtCie2006}, who made comparisons between many grid codes and two SPH codes. They found reasonable agreement in disc structures around embedded protoplanets, though the SPH calculations under-resolved some disc features. This difference is mitigated to an extent in our models by the increased number of particles, more than 6 times greater than the number used in these previous comparisons. Our migration rates obtained using non-self-gravitating locally-isothermal calculations are in good agreement with the analytic model of \cite{TanTakWar2002}, and  so also with many previous similar numerical models which show similar agreement to the analytic model.

\subsection{Self-gravity}
We introduced disc self-gravity to examine its impact upon migration. \cite{PieHur2005} examined the inclusion of disc self-gravity analytically. They broke down the analysis into two competing influences, that of the circumstellar disc-planet interaction, and the circumstellar disc-circumplanetary disc interaction (gas self-gravity). All of our SPH models include the first of these two interactions, which \cite{PieHur2005} found to accelerate Type I migration. So by introducing self-gravity, we mean the addition of the second interaction. \citeauthor{PieHur2005} found this to asymmetrically shift the outer and inner Lindblad resonances such that the migration rate reduces. More recently, numerical models were performed by \cite{BarMas2008} to further explore the impact of including self-gravity. They ran models to explore the impact of both interactions discussed here, and found, in agreement with \cite{PieHur2005} and the previous numerical models of \cite{NelBen2003}, that the disc-disc interaction reduced the Type I migration rates. The consistent, though small, increases in migration timescales found for the SPH calculations including self-gravity (Fig.~\ref{fig:typecvsb}) are in agreement with these previous results. Whilst the effect is small in our models, in discs with masses several times the minimum mass solar nebula self-gravity might slow migration by a more considerable factor \citep{BarMas2008}.

\subsection{Radiation hydrodynamics}
\begin{figure}
\centering
\includegraphics[width=0.9 \columnwidth]{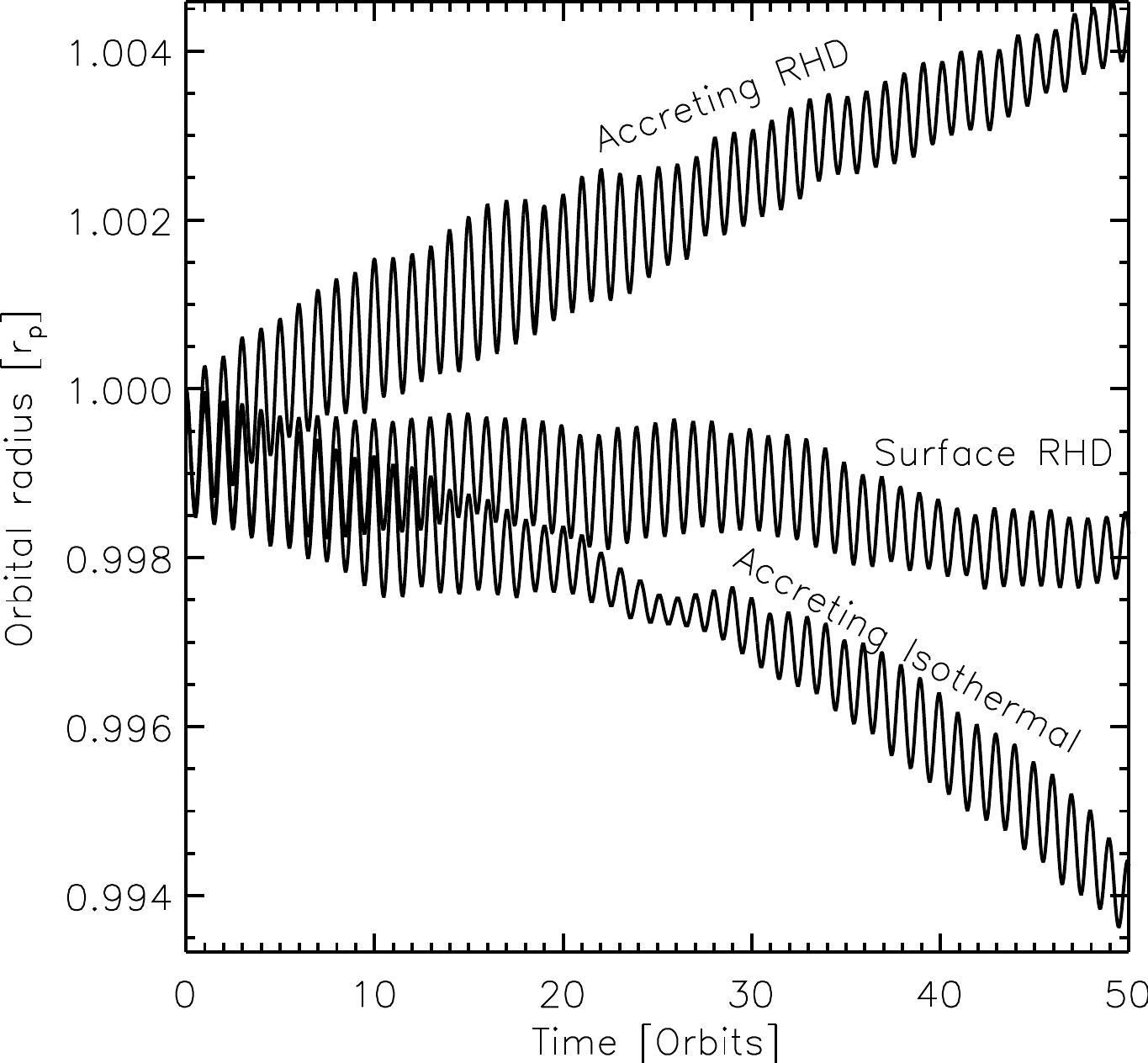}
\caption{The orbital evolution of a 10 \earthmass \ protoplanet modelled three ways, twice as an Accreting sink particle and once as a point mass with surface. The Accreting sink models are differentiated by the chosen equation of state, one locally-isothermal (marked isothermal above), and one using radiation hydrodynamics (marked RHD) which migrates outwards. The point mass with surface is modelled using RHD, and shows a slow but inwards migration.}
\label{fig:outwards}
\end{figure}

Introducing radiative transfer has allowed us to investigate migration beyond the more common locally-isothermal models. Previous studies concerned with the impact of non-isothermality, have found that the inclusion of more complicated thermodynamics leads to increased migration timescales \citep{JanSas2005}, and even the reversal of the Type I migration direction \citep{PaaMel2006, PaaMel2008, KleCri2008, KleBitKla2009}. We are also able to obtain outward migration for a 10 \earthmass \ protoplanet when modelling it with an Accreting sink particle of radius 0.03 \rhill \ with radiation hydrodynamics in an interstellar grain opacity disc, as shown in Fig.~\ref{fig:outwards}.


However, once we progress to using our point mass with surface model, which is the most realistic case, we find no cases of outward migration. The 10 \earthmass \ case modelled in this way, with a surface radius of 0.03~\rhill \  and using interstellar grain opacities, is shown in Fig.~\ref{fig:outwards} for comparison with the Accreting sink model. \cite{KleBitKla2009} find outward migration to result from a density peak in the planet's immediate vicinity, just ahead of it,  which pulls it around in its orbit, increasing its angular momentum and so causing it to migrate outwards. To test the reliability of this result they experiment with different resolutions and various gravitational softening lengths, the shortest being 0.5 \rhill, to explore their impact on the resulting migration. The resulting outward migration was found to be invariant to these changes. The migration rates are based on calculations of torques acting on planet at a fixed orbital radius of 5.2 AU, where the torques are scaled depending upon their distance from the protoplanet, $d$, according to a fraction obtained from the tapering function

\begin{equation}
f_b (d) = \left[ \exp \left( - \frac{d/R_{H} - b}{b/10} \right) + 1 \right]^{-1}.
\label{eq:kley2009}
\end{equation}


\noindent This function is such that at $b$, the torque cut-off radius in units of \rhill, the torques are halved (where they typically use $b = 0.8$). This tapering is applied to avoid large and noisy torques from the poorly resolved Hill radius where the resolution is just 3.3 gridcells/\rhill. Changing the value of $b$ to 0.6 altered their total torques by 10 per cent in locally-isothermal models, and 30 per cent in the radiation hydrodynamics models, but the resolution limits their ability to usefully reduce the value of $b$ further. In Fig.~\ref{fig:torquemap}, where our underlying model resolution is $\approx 20$ times higher than that of \cite{KleBitKla2009}, it can be seen that the greatest torques are acting from within the Roche lobe. In fact the peak torques are acting from a radius of $\approx 0.9$ \rhill \ in the radiation hydrodynamics case where gas is still passing through the Hill sphere, and able to carry away angular momentum. The ability of the SPH model to resolve these torques clearly, removes the need to use a tapering function of the kind above, which otherwise would have scaled these torques down to 73 per cent of their true values. Furthermore, using our planet surface treatment yields a self-consistent envelope around a 20 \earthmass \ protoplanet with densities that are an order of magnitude higher than those obtained using gravitational softening in \citeauthor{KleBitKla2009}'s otherwise similar disc model. Such a change in density is significant given the source of the positive torques suggested by \citeauthor{KleBitKla2009}. Given the improved resolution and realism of the planet treatment employed in our SPH models, our results suggest that outward migration is not an inevitable consequence of introducing more complex thermodynamics.

We find that introducing radiative transfer to our most realistic models of low mass protoplanets in high opacity discs can slow migration rates by up to an order of magnitude. The reduction of Type~I migration rates may be a result of enhanced corotation torques due to entropy gradients across the co-orbital region \citep{PaaPap2008, BarMas2008, KleCri2008, KleBitKla2009, PaaBarCriKle2010}. For high mass protoplanets, those capable of forming circumplanetary discs \citep{AylBat2009b}, the dominance of gravity over thermal effects leaves the migration rates very little altered upon the introduction of radiative transfer when compared with locally-isothermal conditions. This competition between thermal and gravitational influences was also found to play an important role in determining accretion rates onto growing protoplanets by \cite{AylBat2009}. As with migration, the accretion rates in RHD models were found to be most dependent upon the grain opacity for low mass protoplanets, where the gravitational forces at work are smaller. \cite{AylBat2009} found that a 33 \earthmass \ protoplanet, in an interstellar grain opacity disc, is expected to double in mass in $\approx 5 \times 10^{4}$ years. We find the migration timescale for the same protoplanet to be $\approx 5 \times 10^{5}$ years. Under the same conditions a 100 \earthmass \ protoplanet will double in mass in $\approx 9 \times 10^{3}$ years, and has a migration timescale of $\approx 2 \times 10^{4}$ years. Whilst a 333 \earthmass \ protoplanet doubles in mass in $\approx 4 \times 10^{4}$ years and migrates into its central star in $\approx 10^{4}$ years. These results suggest that is should be common for a protoplanet embedded in an interstellar grain opacity disc to reach a Jupiter mass. However, to reach larger masses may require higher disc masses or some additional process to either slow a protoplanet's migration, or accelerate its accretion.

In the low mass regime the increase in migration timescales that we find using RHD may help remedy the problem of proto-giant core depletion that was identified by \cite{AliMorBen2004}, \cite{AliMorBenWin2005}, \cite{IdaLin2008} and \cite{MorAliBen2009}. \citeauthor{IdaLin2008} showed that the efficiency of Type I migration must be reduced by at least an order of magnitude to enable the formation of a population of planets that match current observed prevalences. We find that the inclusion of radiative transfer, in conjunction with a realistic treatment of accretion on to the protoplanet can provide just such a reduction in the low mass regime.

As was discussed in the results section, in the Type II regime the introduction of radiative transfer has two significant effects which have opposing impacts on migration rates. The reduction in the peak density of the spiral arms due to heating results in lower torques from these features, slowing migration, whilst the poorer evacuation of the disc gap leads to a migration process that is intermediate between Types I and II. In all our high mass cases which are modelled with either Accreting sinks, or with surfaces, in locally-isothermal or radiation hydrodynamics models, the rate of migration is considerably faster than the discs viscous migration timescale. Only using the unrealistic Killing sinks do our Type II rates approach the analytic model for Type II migration of \cite{Ward1997}, most likely because they can most rapidly clear a gap of material. Some component of the rapid high mass migration rates found here may result from the models starting with an unperturbed disc, which over 50 orbits fails to establish a gap of the depth that might be expected if the planet had evolved to its initial mass in situ. However, radiative transfer might well be expected to hinder the transition from Type I to the slower Type II migration due to increased thermal pressure that acts to oppose gap formation, as seen in our models.

\section{Summary and Conclusions}

We have conducted models of protoplanet migration with a range of physics and a raft of protoplanet treatments. Beginning with locally-isothermal, non-self gravitating discs, we incrementally added physics to reach self-gravitating radiation hydrodynamic calculations. We progressed from protoplanets modelled by crude sink particles, to those modelled by gravitating point masses with planetary surfaces. The final models we produce are the most self-consistent to date, and turn up some interesting results. In particular,

\begin{itemize}
\item We establish that the `inertial mass problem', suffered by protoplanets that develop circumplanetary discs in models without self-gravity, is an effect of order 10 per cent for minimum mass models.

\item We find that Type I migration rates, those that affect low mass planets (10-33 \earthmass), can be slowed by up to an order of magnitude by the inclusion of radiative transfer in protoplanetary discs with interstellar grain opacities. Such a reduction helps to justify the arbitrary reductions in the Type I migration rate used in planet synthesis models \citep{AliMorBen2004, AliMorBenWin2005, IdaLin2008, MorAliBen2009} where such a reduction is required to ensure planets survive beyond the disc lifetime.

\item The Type I migration timescale that we obtain for a 10 \earthmass \ protoplanet, modelled using RHD in an interstellar grain opacity disc, is of similar order to its mass doubling time, as found in \cite{AylBat2009}, whilst for a 33 \earthmass \ protoplanet the migration timescale is an order of magnitude longer than the mass doubling time. Similarly for a 100 \earthmass \ protoplanet the mass doubling time is shorter than the migration timescale, but for such high mass protoplanets the migration timescale is largely unchanged between locally-isothermal and RHD models. These accretion and migration timescales suggest that a protoplanet embedded in an interstellar grain opacity disc might grow to a high mass within its available migration time.

\item We find that the introduction of radiative transfer to a model of a 10 \earthmass \ protoplanet, represented by an Accreting sink particle embedded in an interstellar grain opacity disc, causes it to migrate outwards. However the inclusion of radiative transfer does not lead to outward migration when we employ a well resolved and more realistic model of the protoplanet with a surface.
\end{itemize}


%

%

\section*{Acknowledgments}

We would like to thank the referee for his perceptive comments, and useful suggestions. The calculations reported here were performed using the University of Exeter's SGI Altix ICE 8200 supercomputer. Many visualisations were produced using SPLASH \citep{splash}, a visualisation tool for SPH that is publicly available at http://www.astro.ex.ac.uk/people/dprice/splash. MRB is grateful for the support of a Philip Leverhulme Prize and a EURYI Award which also supported BAA. This work, conducted as part of the award ``The formation of stars and planets: Radiation hydrodynamical and magnetohydrodynamical simulations"  made under the European Heads of Research Councils and European Science Foundation EURYI (European Young Investigator) Awards scheme, was supported by funds from the Participating Organisations of EURYI and the EC Sixth Framework Programme.

\bibliography{paper.bib}

\end{document}